\documentstyle[psfig,onecolumn]{mn}

\newif\ifAMStwofonts

\ifoldfss
  \ifCUPmtlplainloaded \else
    \NewTextAlphabet{textbfit} {cmbxti10} {}
    \NewTextAlphabet{textbfss} {cmssbx10} {}
    \NewMathAlphabet{mathbfit} {cmbxti10} {} 
    \NewMathAlphabet{mathbfss} {cmssbx10} {} 
  \fi
  \ifAMStwofonts
    \ifCUPmtlplainloaded \else
      \NewSymbolFont{upmath} {eurm10}
      \NewSymbolFont{AMSa} {msam10}
      \NewMathSymbol{\upi}     {0}{upmath}{19}
      \NewMathSymbol{\umu}     {0}{upmath}{16}
      \NewMathSymbol{\upartial}{0}{upmath}{40}
      \NewMathSymbol{\leqslant}{3}{AMSa}{36}
      \NewMathSymbol{\geqslant}{3}{AMSa}{3E}

      \let\geq=\geqslant 
    \fi
  \fi
\fi 

\ifnfssone
  \newmathalphabet{\mathit}
  \addtoversion{normal}{\mathit}{cmr}{m}{it}
  \addtoversion{bold}{\mathit}{cmr}{bx}{it}
  \newmathalphabet{\mathbfit} 
  \addtoversion{normal}{\mathbfit}{cmr}{bx}{it}
  \addtoversion{bold}{\mathbfit}{cmr}{bx}{it}
  \newmathalphabet{\mathbfss} 
  \addtoversion{normal}{\mathbfss}{cmss}{bx}{n}
  \addtoversion{bold}{\mathbfss}{cmss}{bx}{n}
  \ifAMStwofonts
    \ifCUPmtlplainloaded \else
      \UseAMStwoboldmath
      \makeatletter
      \new@mathgroup\upmath@group
      \define@mathgroup\mv@normal\upmath@group{eur}{m}{n}
      \define@mathgroup\mv@bold\upmath@group{eur}{b}{n}
      \edef\UPM{\hexnumber\upmath@group}
      \new@mathgroup\amsa@group
      \define@mathgroup\mv@normal\amsa@group{msa}{m}{n}
      \define@mathgroup\mv@bold\amsa@group{msa}{m}{n}
      \edef\AMSa{\hexnumber\amsa@group}
      \makeatother
      \mathchardef\upi="0\UPM19
      \mathchardef\umu="0\UPM16
      \mathchardef\upartial="0\UPM40
      \mathchardef\leqslant="3\AMSa36
      \mathchardef\geqslant="3\AMSa3E

      \let\geq=\geqslant 
    \fi
  \fi
\fi 

\ifnfsstwo
  \DeclareMathAlphabet{\mathbfit}{OT1}{cmr}{bx}{it}
  \SetMathAlphabet\mathbfit{bold}{OT1}{cmr}{bx}{it}
  \DeclareMathAlphabet{\mathbfss}{OT1}{cmss}{bx}{n}
  \SetMathAlphabet\mathbfss{bold}{OT1}{cmss}{bx}{n}
  \ifAMStwofonts
    \ifCUPmtlplainloaded \else
      \DeclareSymbolFont{UPM}{U}{eur}{m}{n}
      \SetSymbolFont{UPM}{bold}{U}{eur}{b}{n}
      \DeclareSymbolFont{AMSa}{U}{msa}{m}{n}
      \DeclareMathSymbol{\upi}{0}{UPM}{"19}
      \DeclareMathSymbol{\umu}{0}{UPM}{"16}
      \DeclareMathSymbol{\upartial}{0}{UPM}{"40}
      \DeclareMathSymbol{\leqslant}{3}{AMSa}{"36}
      \DeclareMathSymbol{\geqslant}{3}{AMSa}{"3E}

      \let\geq=\geqslant 
    \fi
  \fi
\fi 

\ifCUPmtlplainloaded \else
  \ifAMStwofonts \else 
    \def\upi{\pi}
    \def\umu{\mu}
    \def\upartial{\partial}
  \fi
\fi

\title{Multiple mode gravitational wave detection with a spherical antenna}

\author[J. A. Lobo and M. A. Serrano]
       {J. A. Lobo and M. A. Serrano \\
        Departament de F\'\i sica Fonamental, Universitat de Barcelona \\
	Diagonal 647, E-08028 Barcelona, Spain}
\date{31 March 1999}

\pagerange{\pageref{firstpage}--\pageref{lastpage}}
\pubyear{1999}

\begin{document}

\maketitle

\label{firstpage}

\begin{abstract}
Apart from omnidirectional, a solid elastic sphere is a natural multimode
and multifrequency device for the detection of Gravitational Waves (GW).
Motion sensing in a spherical GW detector thus requires a multiple set of
transducers attached to it at suitable locations. {\it Resonant\/} transducers
exert a significant back action on the larger sphere, and as a consequence
the {\it joint dynamics\/} of the entire system has to be properly understood
before reliable conclusions can be drawn from its readout. In this paper, we
present and develop a mathematical formalism to analyse such dynamics, which
generalises and enhances currently existing ones, and which clarifies their
actual range of validity, thereby shedding light into the physics of the
detector. In addition, the new formalism has enabled us to discover a new
resonator layout (we call it {\sl PHC\/}) which only requires five resonators
per quadrupole mode sensed, and has {\it mode channels\/}, i.e., linear
combinations of the transducers' readouts which are directly proportional to
the GW amplitudes. The {\it perturbative\/} nature of our proposed approach
makes it also very well adapted to systematically assess the consequences of
small mistunings in the device parameters by robust analytic methods. These
prove to be very powerful, not only theoretically but also practically:
confrontation of our model's predictions with the experimental data of the
{\sl LSU\/} prototype detector {\sl TIGA\/} reveals an agreement between
both consistently reaching the {\it fourth\/} decimal place.
\end{abstract}


\section{Introduction}
\label{sec:intro}

The idea of using a solid elastic sphere as a gravitational wave (GW)
antenna is almost as old as that of using cylindrical bars: as far back
as 1971 Forward published a paper \cite{fo71} in which he assessed some
of the potentialities offered by a spherical solid for that purpose. It
was however Weber's ongoing philosophy and practice of using bars which
eventually prevailed and developped up to the present date, with the highly
sophisticated and sensitive ultracryogenic systems currently in operation
---see \cite{amaldi} and \cite{gr14} for reviews and bibliography. With few
exceptions~\cite{ad75,wp77}, spherical detectors fell into oblivion for
years, but interest in them strongly re-emerged in the early 1990's, and
an important number of research articles have been published since which
address a wide variety of problems in GW spherical detector science. At
the same time, international collaboration has intensified, and prospects
for the actual construction of large spherical GW observatories (in the
range of $\sim$100 tons) are being currently considered in several countries
\footnote{
There are collaborations in Brazil, Holland, Italy and Spain.},
even in a variant {\it hollow\/} shape \cite{vega}.

A spherical antenna is obviously omnidirectional but, most important, it
is also a natural {\it multimode\/} device, i.e., when suitably monitored,
it can generate information on all the GW amplitudes and incidence direction
\cite{nadja}, a capability which possesses no other {\it individual\/} GW
detector, whether resonant or interferometric \cite{dt}. Furthermore, a
spherical antenna could also reveal the eventual existence of {\it monopole\/}
gravitational radiation, or set thresholds on it \cite{maura}. The
theoretical explanation of these facts is to be found in the unique
matching between the GW amplitude structure and that of the sphere
oscillation eigenmodes: a general {\it metric\/} GW generates a
{\it tidal\/} field of forces in an elastic body which is given in terms
of the ``electric'' components $R_{0i0j}(t)$ of the Riemann tensor at its
centre of mass by the following formula \cite{lobo}:

\begin{equation}
    {\bf f}_{\rm GW}({\bf x},t)\ \ \ =
    \sum_{\stackrel{\scriptstyle l=0\ {\rm and}\ 2}{m=-l,...,l}}\,
    {\bf f}^{(lm)}({\bf x})\,g^{(lm)}(t)    \label{1.1}
\end{equation}
where ${\bf f}^{(lm)}({\bf x})$ are ``tidal form factors'', while
$g^{(lm)}(t)$ are specific linear combinations of the Riemann tensor
components $R_{0i0j}(t)$ which carry all the {\it dynamical\/} information
on the GW's monopole ($l\/$\,=\,0) and quadrupole ($l\/$\,=\,2) amplitudes.
It is precisely these amplitudes which a GW detector is aimed to measure.

On the other hand, a free elastic sphere has two families of oscillation
eigenmodes, so called {\it toroidal\/} and {\it spheroidal\/} modes, and
modes within either family group into ascending series of $l\/$-pole
harmonics, each of whose frequencies is (2$l\/$+1)-fold degenerate ---see
\cite{lobo} for full details. It so happens that {\it only\/} monopole
and/or quadrupole spheroidal modes can possibly be excited by an incoming
{\it metric\/} GW \cite{bian}, and their GW driven amplitudes are directly
proportional to the wave amplitudes $g^{(lm)}(t)$ of equation (\ref{1.1}).
It is this very fact which makes of the spherical detector such a natural
one for GW observations \cite{lobo}. In addition, a spherical antenna has
a significantly higher absorption {\it cross section\/} than a cylinder of
like fundamental frequency, and also presents good sensitivity at the
{\it second\/} quadrupole harmonic \cite{clo}.

In order to monitor the GW induced deformations of the sphere {\it motion
sensors\/} are required. In cylindrical bars, current state of the art
technology is based upon {\it resonant transducers\/} \cite{pia,hamil}.
A resonant transducer consists in a small (compared to the bar) mechanical
device possessing a resonance frequency accurately tuned to that of
the cylinder. This {\it frequency matching\/} causes back-and-forth
{\it resonant energy transfer\/} between the two bodies (bar and resonator),
which results in turn in {\it mechanically amplified\/} oscillations of the
smaller resonator. The philosophy of using resonators for motion sensing is
directly transplantable to a spherical detector ---only a {\it multiple\/}
set rather than a single resonator is required if its potential capabilities
as a multimode system are to be exploited to satisfaction.

A most immediate question in a multiple motion sensor system is:
{\it where\/} should the sensors be? The answer to this basic question
naturally depends on design and purpose criteria. Merkowitz and Johnson
(M\&J) made a very appealing proposal consisting in a set of 6 identical
resonators coupling to the {\it radial\/} motions of the sphere's surface,
and occupying the positions of the centres of the 6 non-parallel pentagonal
faces of a truncated icosahedron \cite{jm93,jm95}. One of the most remarkable
properties of such layout is that there exist 5 linear combinations of the
resonators' readouts which are directly proportional to the 5 quadrupole
GW amplitudes $g^{(2m)}(t)$ of equation (\ref{1.1}). M\&J call these
combinations {\it mode channels\/}, and they therefore play a fundamental
role in GW signal deconvolution in a real, {\it noisy\/} system
\cite{m98,lms}. In addition, a reduced scale prototype antenna ---called
{\sl TIGA\/}, for {\sl T\/}runcated {\sl I\/}cosahedron {\sl G\/}ravitational
{\sl A}ntenna--- was constructed at Lousiana State University, and its
working experimentally put to test~\cite{phd}. The remarkable success of
this experiment in almost every detail \cite{jm96,jm97,jm98} stands as
a vivid proof of the practical feasibility of a spherical GW detector
\cite{sfera}.

Despite its success, the theoretical model proposed by M\&J to describe the
system dynamics is based upon a simplifying assumption that the resonators
{\it only\/} couple to to the quadrupole vibration modes of the sphere
\cite{jm93,jm95}. While this is seen {\it a posteriori\/} of experimental
measurements to be a very good approximation \cite{phd,jm97}, a deeper
{\it physical\/} reason which explains {\it why\/} this happens is missing
so far. The original motivation for the research we present in this article
was to develop a new and more general approach for the analysis of the
resonator problem, very much in the spirit of the methodology and results
of reference \cite{lobo}; this, we thought, would not only provide the
necessary tools for a rigorous analysis of the system dynamics, but also
contribute to improve our understanding of the physics of the spherical
GW detector.

Pursuing this programme, we succeeded in setting up and solving the equations
of motion for the coupled system of sphere plus resonators. The most important
characteristic of our solution is that it is expressible as a
{\it perturbative series expansion in ascending powers of the small parameter
$\eta^{1/2}$\/}, where $\eta\/$ is the ratio between the average resonator's
mass and the sphere's mass. The dominant (lowest) order terms in this
expansion appear to exactly reproduce Merkowitz and Johnson's equations
\cite{jm95}, whence a quantitative assessment of their degree of accuracy,
as well as of the range of validity of their underlying hypotheses obtains;
if further precision is required then a well defined procedure of going to
next (higher) order terms is unambiguously prescribed by the system equations.

Beyond this, though, the simple and elegant algebra which emerges out of
the general scheme has enabled us to explore different resonator layouts,
alternative to the unique {\sl TIGA\/} of M\&J. In particular we have found
one \cite{ls,lsc} requiring 5 rather than 6 resonators per quadrupole mode
sensed and possessing the remarkable property that {\it mode channels\/} can
be constructed from the system readouts, i.e., five linear combinations of the
latter which are directly proportional to the five quadrupole GW amplitudes.
We have called this distribution {\sl PHC\/} ---see below for full details.

The intrinsically perturbative nature of our approach makes it also
particularly well adapted to assess the consequences of small defects in the
system structure, such as for example symmetry breaking due to suspension
attachments, small resonator mistunings and mislocations, etc.\ We have
applied them with outstanding success to account for the reported frequency
measurements of the {\sl LSU TIGA\/} prototype \cite{phd}, which was
diametrically drilled for suspension purposes; in particular, discrepancies
between measured and calculated values (generally affecting only the
{\it fourth\/} decimal place) are precisely of the theoretically predicted
order of magnitude.

We have also applied our methods to analyse the stability of the spherical
detector to several mistuned parameters, with the result that it is not very
sensitive to small construction errors. This conforms again to experimental
reports \cite{jm98}, but has the advantage that the argument depends on
{\it analytic\/} mathematical work rather than on computer simulated
procedures, the only ones available to this date to our knowledge ---see
e.g.\ \cite{jm98} or \cite{ts}.

The paper is structured as follows: in section~\ref{sec:GE} we present the
main physical hypotheses of the model, and the general equations of motion.
In section~\ref{sec:gff} we set up a Green function approach to solve those
equations, and in section~\ref{sec:srgw} we apply it to assess the system
response to both monopole and quadrupole GW signals. In section~\ref{sec:PHC}
we describe in detail the {\sl PHC\/} layout, including its frequency spectrum
and {\it mode channels\/}. Section~\ref{sec:hs} contains a few brief
considerations on the system response to a hammer stroke calibration signal,
and finally in section~\ref{sec:symdef} we study how the different parameter
mistunings affect the detector's behaviour. The paper closes with a summary
of conclusions, and three appendices where the heavier mathematical details
are made precise for the interested reader.

\section{General equations}  \label{sec:GE}

\begin{figure}
\label{fig1}
\psfig{file=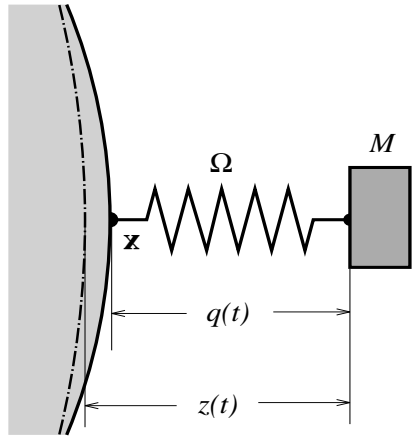,height=17cm,width=12cm,rheight=6.8cm,bbllx=-3cm,bblly=-7.4cm,bburx=14.2cm,bbury=19cm}
\caption{Schematic diagramme of the coupling model between a solid sphere
and a resonator. The notation is that in the text, but subindices have been
dropped for clarity. The dashed-dotted arc line on the left indicates the
position of the {\it undeformed\/} sphere's surface, and the solid arc its
{\it actual\/} position.}
\end{figure}

With minor improvements, we shall use the notation of references \cite{lobo}
and \cite{ls}, some of which is now briefly recalled. We consider a solid
sphere of mass $\cal M\/$, radius $R\/$, (uniform) density $\varrho\/$,
and elastic Lam\'e coefficients $\lambda$ and $\mu\/$, endowed with
a set of $J\/$ resonators of masses $M_a\/$ and resonance frequencies
$\Omega_a\/$ ($a\/$\,=\,1,\ldots,$J\/$), respectively. We shall model the
latter as {\it point masses\/} attached to one end of a linear spring,
whose other end is rigidly linked to the sphere at locations~${\bf x}_a\/$
---see Figure \ref{fig1}. The system degrees of freedom are given by the
{\it field\/} of elastic displacements ${\bf u}({\bf x},t)$ of the sphere
plus the {\it discrete\/} set of resonator spring deformations $z_a(t)$;
equations of motion need to be written down for them, of course, and this
is our next concern in this section.

We shall assume that the resonators only move radially, and also that
Classical Elasticity theory \cite{ll70} is sufficiently accurate for our
purposes\footnote{
We clearly do not expect relativistic motions in extremely small displacements
at typical frequencies in the range of 1 kHz.}.
In these circumstances we have \cite{ls}

\begin{eqnarray}
    \varrho\,\frac{\partial^2 {\bf u}}{\partial t^2} & = & \mu\nabla^2 {\bf u}
    + (\lambda+\mu)\,\nabla(\nabla{\bf\cdot}{\bf u}) + {\bf f}({\bf x},t)
    \label{2.1.a}  \\*[0.7 em]
    \ddot{z}_a(t) & = & -\Omega_a^2\,
    \left[z_a(t)-u_a(t)\right]+\xi_a^{\rm external}(t)\ , \qquad a=1,\ldots,J
    \label{2.1.b}
\end{eqnarray}
where ${\bf n}_a\/$\,$\equiv$\,${\bf x}_a/R\/$ is the outward pointing normal
at the the $a\/$-th resonator's attachment point, and

\begin{equation}
  u_a(t)\equiv{\bf n}_a\!\cdot\!{\bf u}({\bf x}_a,t)\ ,\qquad
  a=1,\ldots,J    \label{3.8}
\end{equation}
is the {\it radial\/} deformation of the sphere's surface at ${\bf x}_a\/$.
A dot (\,$\dot{}$\,) is an abbreviation for time derivative. The term in
square brackets in (\ref{2.1.b}) is thus the spring deformation ---$q(t)$ in
Figure \ref{fig1}.

${\bf f}({\bf x},t)$ in the rhs of (\ref{2.1.a}) contains the
{\it density\/} of all {\it non-internal\/} forces acting on the sphere,
which we expediently split into a component due the resonators' {\it back
action\/} and an external action {\it proper\/}, which can be a GW signal,
a calibration signal, etc. Thus

\begin{equation}
  {\bf f}({\bf x},t) = {\bf f}_{\rm resonators}({\bf x},t) +
  {\bf f}_{\rm external}({\bf x},t)    \label{2.2}
\end{equation}

Finally, $\xi_a^{\rm external}(t)$ in the rhs of (\ref{2.1.b}) is the
force per unit mass (acceleration) acting on the $a\/$-th resonator due
to {\it external\/} agents.

Since we are making the hypothesis that the resonators are point masses the
following holds:

\begin{equation}
    {\bf f}_{\rm resonators}({\bf x},t) =
    \sum_{a=1}^J M_a\Omega_a^2\,\left[\,z_a(t)-u_a(t)\right]\,
    \delta^{(3)}({\bf x}-{\bf x}_a)\,{\bf n}_a
    \label{2.3}
\end{equation}
where $\delta^{(3)}\/$ is the three dimensional Dirac density function.

The {\it external\/} forces we shall be considering in this paper will be
{\it gravitational wave\/} signals, and also a simple calibration signal,
a perpendicular {\it hammer stroke\/}. GW driving terms, we recall
from~(\ref{1.1}), can be written

\begin{equation}
   {\bf f}_{\rm GW}({\bf x},t) = {\bf f}^{(00)}({\bf x})\,g^{(00)}(t)\ +\ 
    \sum_{m=-2}^2\,{\bf f}^{(2m)}({\bf x})\,g^{(2m)}(t)    \label{2.4}
\end{equation}
for a general {\it metric\/} wave ---see \cite{lobo} for explicit
formulas and technical details. While the spatial coefficients
${\bf f}^{(lm)}({\bf x})$ are pure {\it form factors\/} associated to
the {\it tidal\/} character of a GW excitation, it is the time dependent
factors $g^{(lm)}(t)$ which carry the specific information on the
incoming GW. The purpose of a GW detector is to determine the latter
coefficients on the basis of suitable measurements.

If a GW sweeps the observatory then the resonators themselves will also be
affected, of course. They will be driven, relative to the sphere's centre,
by a tidal acceleration which, since they only move radially, is given by

\begin{equation}
   \xi_a^{\rm GW}(t) = c^2\,R_{0i0j}(t)\,x_{a,i}n_{a,j}\ ,
   \qquad a=1,\ldots,J      \label{c.1}
\end{equation}
where $R_{0i0j}(t)$ are the ``electric'' components of the GW Riemann tensor
at the centre of the sphere. These can be easily manipulated to give\footnote{
$Y_{lm}({\bf n})$ are spherical harmonics \protect\cite{Ed60} ---see also the
multipole expansion of $R_{0i0j}(t)$ in reference \cite{lobo}.}

\begin{equation}
   \xi_a^{\rm GW}(t) = R\,
   \sum_{\stackrel{\scriptstyle l=0\ {\rm and}\ 2}{m=-l,...,l}}\,
   Y_{lm}({\bf n}_a)\,g^{(lm)}(t)\ ,\qquad a=1,\ldots,J   \label{c.4}
\end{equation}
where $R\/$ is the sphere's radius.

We shall also be later considering in this paper the response of the
system to a particular {\it calibration\/} signal, consisting in a hammer
stroke with intensity ${\bf f}_0$, delivered perpendicularly to the
sphere's surface at point ${\bf x}_0$:

\begin{equation}
   {\bf f}_{\rm stroke}({\bf x},t) = {\bf f}_0\,
   \delta^{(3)}({\bf x}-{\bf x}_0)\,\delta(t)   \label{2.5}
\end{equation}
which we have modeled as an impulsive force in both space and time
variables. Unlike GW tides, a hammer stroke will be applied on the sphere's
surface, so it has no {\it direct\/} effect on the resonators. In other words,

\begin{equation}
   \xi_a^{\rm stroke}(t) = 0\ ,\qquad a=1,\ldots,J    \label{c.5}
\end{equation}

Our fundamental equations thus finally read:

\begin{eqnarray}
    \varrho \frac{\partial^2 {\bf u}}{\partial t^2} & = & \mu\nabla^2 {\bf u}
    + (\lambda+\mu)\,\nabla(\nabla{\bf\cdot}{\bf u}) +
    \nonumber \\
    & & \sum_{b=1}^J M_b\Omega_b^2\,\left[z_b(t)-u_b(t)\right]\,
    \delta^{(3)}({\bf x}-{\bf x}_b)\,{\bf n}_b
    + {\bf f}_{\rm external}({\bf x},t)    \label{2.6.a}  \\*[0.7 em]
    \ddot{z}_a(t) & = & -\Omega_a^2\,
    \left[z_a(t)-u_a(t)\right] + \xi_a^{\rm external}(t)\ ,
    \qquad a=1,\ldots,J    \label{2.6.b}
\end{eqnarray}
where ${\bf f}_{\rm external}({\bf x},t)$ will be given by either (\ref{2.4})
or (\ref{2.5}), as the case may be. Likewise, $\xi_a^{\rm external}(t)$ will
be given by (\ref{c.4}) or (\ref{c.5}), respectively. The remainder of this
paper will be concerned with finding solutions to the system of coupled
differential equations (\ref{2.6.a}) and (\ref{2.6.b}), and with their
meaning and consequences.

\section{Green function formalism}  \label{sec:gff}

In order to solve equations~(\ref{2.6.a})-(\ref{2.6.b}) we shall resort to
Green function formalism. The essentials of this procedure in the context
of the present problem can be found in detail in reference~\cite{lobo};
more specific technicalities are given in appendix~\ref{app:a}.

By means of such formalism equations~(\ref{2.6.a})-(\ref{2.6.b}) become
the following integro-differential system:

\begin{eqnarray}
  u_a(t) & = & u_a^{\rm external}(t) + \sum_{b=1}^J\,\eta_b\,\int_0^t
  K_{ab}(t-t')\,\left[\,z_b(t')-u_b(t')\right]\,dt'  \label{3.7.a}\\[1 ex]
  \ddot{z}_a(t) & = & \xi_a^{\rm external}(t)
  -\Omega_a^2\,\left[\,z_a(t)-u_a(t)\right]\ , \qquad a=1,\ldots,J
  \label{3.7.b}
\end{eqnarray}
where $u_a^{\rm external}(t)$\,$\equiv$\,
${\bf n}_a\!\cdot\!{\bf u}^{\rm external}({\bf x}_a,t)$, and
${\bf u}^{\rm external}({\bf x},t)$ is the {\it bare\/} (i.e., without
attached resonators) sphere's response to the external forces
${\bf f}_{\rm external}({\bf x},t)$ in the rhs of~(\ref{2.6.a}).
$K_{ab}(t)$ is a {\it kernel matrix\/} defined by the following weighted
sum of diadic products of wavefunctions\footnote{
We shall often use the capitalised index $N\/$ to imply the multiple index
$\{nlm\}$ which characterises the sphere's wavefunctions.}:

\begin{equation}
  K_{ab}(t) = \Omega_b^2\,\sum_N\,\omega_N^{-1}\,
  \left[{\bf n}_b\!\cdot\!{\bf u}_N^*({\bf x}_b)\right]
  \left[{\bf n}_a\!\cdot\!{\bf u}_N({\bf x}_a)\right]\,\sin\omega_Nt
  \label{3.10}
\end{equation}

Finally, we have defined the mass ratios of the resonators to the entire
sphere

\begin{equation}
  \eta_b\equiv \frac{M_b}{\cal M}\ ,\qquad b=1,\ldots,J   \label{3.11}
\end{equation}
which will be {\it small parameters\/} in a real device.

Before proceeding further, let us briefly pause for a qualitative inspection
of equations~(\ref{3.7.a}) and~(\ref{3.7.b}). Equation (\ref{3.7.a}) shows
that the sphere's surface deformations $u_a(t)$ are made up of two
contributions: one due to the action of {\it external\/} agents (GWs or
other), contained in $u_a^{\rm external}(t)$, and another one due to coupling
to the resonators. The latter is commanded by the small parameters $\eta_b\/$,
and correlates to {\it all\/} of the sphere's spheroidal eigenmodes through
the kernel matrix $K_{ab}(t)$. This has consequences for GW detectors, for
even though GWs may only couple to quadrupole and monopole\footnote{
Monopole modes only exist in scalar-tensor theories of gravity, such as e.g.
Brans--Dicke \protect\cite{bd61}; General Relativity does not belong in this
category.}
spheroidal modes of the {\it free\/} sphere \cite{lobo,bian},
attachment of resonators causes, as we see, coupling between these and the
other modes of the antenna, and conversely, these modes back-act on the
former. As we shall shortly prove, such effects can be minimised by suitable
{\it tuning\/} of the resonators' frequencies.

\subsection{Laplace transform domain equations}

We now take up the problem of solving these equations. Equation (\ref{3.7.a})
is an integral equation belonging in the general category of Volterra
equations \cite{tricomi}, but the usual iterative solution to it by repeated
substitution of $u_b(t)$ into the kernel integral is not viable here due to
the {\it dynamical\/} contribution of $z_b(t)$, which is in turn governed by
the {\it differential\/} equation~(\ref{3.7.b}).

A better suited method to solve this {\it integro-differential\/} system is
to Laplace-transform it. We denote the Laplace transform of a generic function
of time $f(t)$ with a {\it caret\/} (\,$\hat{}$\,) on its symbol, e.g.,

\begin{equation}
  \hat{f}(s) \equiv \int_0^\infty f(t)\,e^{-st}\,dt	\label{3.12}
\end{equation}
and make the assumption that the system is at rest before an instant of
time, $t\/$\,=\,0, say, or

\begin{equation}
  {\bf u}({\bf x},0)={\bf\dot u}({\bf x},0)=z_a(0)=\dot z_a(0) = 0
  \label{3.14}
\end{equation}

Equations (\ref{3.7.a}) and (\ref{3.7.b}) then adopt the equivalent form

\begin{eqnarray}
    \hat u_a(s) & = & \hat u_a^{\rm external}(s)
    - \,\sum_{b=1}^J \eta_b\,\hat K_{ab}(s)\,
    \left[\hat z_b(s)-\hat u_b(s)\right]
    \label{3.13.a}   \\*[0.7 em]
    s^2\,\hat{z}_a(s) & = & \hat\xi_a^{\rm external}(s) -
    \Omega_a^2\,\left[\hat z_a(s)-\hat u_a(s)\right]\ ,\qquad a=1,\ldots,J
    \label{3.13.b}
\end{eqnarray}
for which use has been made of the {\it convolution theorem\/} for Laplace
transforms\footnote{
This theorem states, it is recalled, that the Laplace transform of the
convolution product of two functions is the arithmetic product of their
respective Laplace transforms.}.
A further simplification is accomplished if we consider that we shall
in practice be only concerned with the {\it measurable\/} quantities

\begin{equation}
  q_a(t)\equiv z_a(t)-u_a(t) \ ,\qquad  a=1,\ldots,J	 \label{3.15}
\end{equation}
representing the resonators' actual elastic deformations ---cf.\ Figure
\ref{fig1}. It is readily seen that these verify the following:

\begin{equation}
  \sum_{b=1}^J \left[\delta_{ab} + \eta_b\,\frac{s^2}{s^2+\Omega_a^2}\,
  \hat K_{ab}(s)\right]\,\hat q_b(s) = -\frac{s^2}{s^2+\Omega_a^2}\,
  \hat u_a^{\rm external}(s) + \frac{\hat\xi_a^{\rm external}(s)}
  {s^2+\Omega_a^2}\ ,\qquad a=1,\ldots,J
  \label{3.16}
\end{equation}

Equations (\ref{3.16}) constitute a significant simplification of the
original problem, as they are a set of just $J\/$ {\it algebraic\/} rather
than integral or differential equations. We must solve them for the unknowns
$\hat q_a(s)$, then perform {\it inverse Laplace transforms\/} to revert to
$q_a(t)$. We do this next.

\section{System response to a Gravitational Wave}
\label{sec:srgw}

Our concern now is the actual system response when it is acted upon by
an incoming GW. We shall calculate it by making a number of simplifying
assumptions, more precisely:

\begin{enumerate}
 \item[\sf i)] The detector is perfectly spherical.
 \item[\sf ii)] The resonators have identical masses and resonance frequencies.
 \item[\sf iii)] The resonators' frequency is accurately matched to one of
		 the sphere's oscillation eigenfrequencies.
\end{enumerate}

As we shall see below (section \ref{sec:symdef}), a real system can be
appropriately treated as one which deviates by definite amounts from this
idealised construct. Therefore detailed knowledge of the ideal system
behaviour is essential for all purposes. This is the justification for the
above simplifications.

The wavefunctions ${\bf u}_{nlm}({\bf x})$ of an elastic sphere can be
found in reference~\cite{lobo} in full detail, and we shall keep the
notation of that paper for them. The Laplace transform of the kernel
matrix~(\ref{3.10}) can thus be expressed as ---see equation (\ref{A3.20})
in appendix~\ref{app:a}:

\begin{equation}
  \hat K_{ab}(s) = \sum_{nl}\,\frac{\Omega_b^2}{s^2+\omega_{nl}^2}\,
   \left|A_{nl}(R)\right|^2\,\frac{2l+1}{4\pi}\,
   P_l({\bf n}_a\!\cdot\!{\bf n}_b) \equiv
   \sum_{nl}\,\frac{\Omega_b^2}{s^2+\omega_{nl}^2}\,\chi_{ab}^{(nl)}
   \label{4.2}
\end{equation}
where the last term simply {\it defines\/} the quantities $\chi_{ab}^{(nl)}$.
Note that the sums here stretch across the {\it entire\/} spectrum of the
solid sphere.

Our next assumption that all the resonators are {\it identical\/} simply
means that

\begin{equation}
  \eta_1=\,\ldots\,=\eta_J\equiv\eta\ ,\qquad
  \Omega_1=\,\ldots\,=\Omega_J\equiv\Omega
  \label{4.5}
\end{equation}

The third hypothesis makes reference to the fundamental idea behind using
resonators, which is to have them tuned to one of the frequencies of the
sphere's spectrum. We express this by

\begin{equation}
  \Omega = \omega_{n_0l_0}	\label{4.6}
\end{equation}
where $\omega_{n_0l_0}$ is a specific and {\it fixed\/} frequency of the
spheroidal spectrum.

In a GW detector it will only make sense to choose $l_0$\,=\,0 or
$l_0$\,=\,2, as only monopole and quadrupole sphere modes couple to the
incoming signal; in practice, $n_0$ will refer to the first or perhaps
second harmonic \cite{clo}. We shall however keep the generic expression
(\ref{4.6}) for the time being in order to encompass all the possibilities
with a unified notation.

Based on the above hypotheses, equation~(\ref{3.16}) can be rewritten in
the form

\begin{equation}
 \sum_{b=1}^J\,\left[\delta_{ab} + \eta\,\sum_{nl}\,
   \frac{\Omega^2s^2}{(s^2+\Omega^2)(s^2+\omega_{nl}^2)}\,\chi_{ab}^{(nl)}
   \right]\,\hat q_b(s) = -\frac{s^2}{s^2+\Omega^2}\,
   \hat u_a^{\rm GW}(s) + \frac{\hat\xi_a^{\rm GW}(s)}
   {s^2+\Omega^2}\ ,\qquad  (\Omega = \omega_{n_0l_0})
   \label{4.8}
\end{equation}
where $\hat\xi_a^{\rm GW}(s)$ is the Laplace transform of~(\ref{c.4}), i.e.,

\begin{equation}
   \hat\xi_a^{\rm GW}(s) = R\,
   \sum_{\stackrel{\scriptstyle l=0\ {\rm and}\ 2}{m=-l,...,l}}\,
   Y_{lm}({\bf n}_a)\,\hat g^{(lm)}(s)\ ,\qquad a=1,\ldots,J
   \label{4.85}
\end{equation}

As mentioned at the end of the previous section, we must now invert the
matrix in the lhs of~(\ref{4.8}), which will give us an expression for
$\hat q_a(s)$, then find the {\it inverse Laplace transform\/} of these
functions to revert back to the time domain. A simple glance at the
equation suffices however to grasp the unsurmountable difficulties of
accomplishing this {\it analytically\/}.

Thankfully, though, a {\it perturbative\/} approach is applicable when
the masses of the resonators are small compared to the mass of the whole
sphere, i.e., when the inequality

\begin{equation}
  \eta\ll 1	\label{4.10}
\end{equation}
holds. We shall henceforth assume that this is the case, as also is with
cylindrical bar resonant transducers. It is shown in appendix~\ref{app:b}
that the perturbative series happens in ascending powers of $\eta^{1/2}$,
rather than $\eta\/$ itself, and that the lowest order contribution has
the form

\begin{equation}
    \hat q_a(s) = \eta^{-1/2}\,\sum_{l,m}\,\hat\Lambda_a^{(lm)}(s;\Omega)\,
    \hat g^{(lm)}(s) + O(0) \ ,\qquad a=1,\ldots,J
    \label{6.8}
\end{equation}
where $O(0)$ stands for terms of order $\eta^0$ or smaller. Here,
$\hat\Lambda_a^{(lm)}(s;\Omega)$ is a {\it transfer function matrix\/}
which relates {\it linearly\/} the system response $\hat q_a(s)$ to the
GW amplitudes $\hat g^{(lm)}(s)$, in the usual sense that $q_a(t)$ is
given by the {\it convolution product\/} of the signal $g^{(lm)}(t)$
with the time domain expression, $\Lambda_a^{(lm)}(t;\Omega)$, of
$\hat\Lambda_a^{(lm)}(s;\Omega)$. The detector is thus seen to act as
a {\it linear filter\/} on the GW signal, whose frequency response is
characterised by the properties of $\hat\Lambda_a^{(lm)}(s;\Omega)$.
More specifically, the filter has a number of characteristic frequencies
which correspond to the {\it imaginary parts of the poles\/} of
$\hat\Lambda_a^{(lm)}(s;\Omega)$. As also shown in appendix~\ref{app:b},
these frequencies are the symmetric pairs

\begin{equation}
  \omega_{a\pm}^2 = \Omega^2\,\left(1\pm\sqrt{\frac{2l+1}{4\pi}}\,
  \left|A_{n_0l_0}(R)\right|\,\zeta_a\,\eta^{1/2}\right) + O(\eta)\ ,
  \qquad a=1,\ldots,J
  \label{5.2}
\end{equation}
where $\zeta_a^2\/$ is the $a\/$-th eigenvalue of the Legendre matrix

\begin{equation}
  P_{l_0}({\bf n}_a\!\cdot\!{\bf n}_b)\ ,\qquad a,b=1,\,\ldots,J
  \label{5.25}
\end{equation}
associated to the multipole ($l_0$) selected for tuning ---see (\ref{4.6}).
These frequency pairs correspond to {\it beats\/}, typical of resonantly
coupled oscillating systems ---we shall find them again in
section~\ref{sec:hs} in a particularly illuminating example.

Another very important fact is also neatly displayed by equation~(\ref{6.8}):
the resonators' motions are {\it mechanically amplified\/} by a factor
$\eta^{-1/2}$ relative to the driving amplitudes $\hat g^{(lm)}(s)$. This
is the counterpart, in our multimode system, of a similar behaviour known
to happen in monomode cylindrical antennas \cite{pia}.

The specific form of the transfer function matrix
$\hat\Lambda_a^{(lm)}(s;\Omega)$ depends on both the selected mode to
tune the resonator frequency $\Omega$, and on the resonator distribution
geometry. We now come to a discussion of these.

\subsection{Monopole gravitational radiation sensing}

General Relativity, as is well known, forbids monopole GW radiation. More
general {\it metric\/} theories, e.g. Brans-Dicke \cite{bd61}, do however
predict this kind of radiation. It appears that a spherical antenna is
potentially sensitive to monopole waves, so it can serve the purpose of
thresholding, or eventually detecting them. This clearly requires that the
resonator set be tuned to a monopole harmonic of the sphere, i.e.,

\begin{equation}
   \Omega = \omega_{n0}\ ,\qquad (l_0=0)	\label{6.9}
\end{equation}
where $n\/$ tags the chosen harmonic ---most likely the first ($n\/$\,=\,1)
in a thinkable device.

Since $P_0(z)$\,$\equiv$\,1 (for all $z\/$) the eigenvalues of
$P_0({\bf n}_a\!\cdot\!{\bf n}_b)$ are, clearly,

\begin{equation}
  \zeta_1^2=J\ ,\qquad \zeta_2^2=\,\ldots\,=\zeta_J^2=0
  \label{6.10}
\end{equation}
for {\it any resonator distribution\/}. The tuned mode frequency thus splits
into a {\it single\/} strongly coupled pair:

\begin{equation}
  \omega_\pm^2 = \Omega^2\,\left(1\pm\sqrt{\frac{J}{4\pi}}\,
  \left|A_{n0}(R)\right|\,\eta^{1/2}\right) + O(\eta)\ ,
  \qquad \Omega=\omega_{n0}
  \label{6.11}
\end{equation}

The $\Lambda$-matrix of equation (\ref{6.8}) is seen to be in
this case

\begin{equation}
  \hat\Lambda_a^{(lm)}(s;\omega_{n0}) = (-1)^J\,\frac{a_{n0}}{\sqrt{J}}\,
  \frac{1}{2}\,\left[\left(s^2+\omega_+^2\right)^{-1} -
  \left(s^2+\omega_-^2\right)^{-1}\right]\,\delta_{l0}\,\delta_{m0}
  \label{6.12}
\end{equation}
whence the system response is

\begin{equation}
  \hat q_a(s) = \eta^{-1/2}\,\frac{(-1)^J}{\sqrt{J}}\,a_{n0}\,
  \frac{1}{2}\,\left[\left(s^2+\omega_+^2\right)^{-1} -
  \left(s^2+\omega_-^2\right)^{-1}\right]\,\hat g^{(00)}(s) + O(0)\ ,
  \qquad a=1,\ldots,J
  \label{6.13}
\end{equation}
{\it regardless of resonator positions\/}. The overlap coefficient
$a_{n0}$ is calculated by means of formulas given in \cite{lobo}, and
has dimensions of length. By way of example, $a_{10}/R\/$\,=\,0.214,
and $a_{20}/R\/$\,=\,$-$0.038 for the first two harmonics.

A few interesting facts are displayed by equation (\ref{6.13}). First, as
we have already stressed, it is seen that if the resonators are tuned to
a monopole {\it detector\/} frequency then only monopole {\it wave
amplitudes\/} couple strongly to the system, even if quadrupole radiation
amplitudes are significantly high at the observation frequencies
$\omega_\pm\/$. Also, the amplitudes $\hat q_a(s)$ are equal for all $a\/$,
as corresponds to the spherical symmetry of monopole sphere's oscillations,
and are proportional to $J^{-1/2}$, a factor we should indeed expect as an
indication that GW {\it energy\/} is evenly distributed amongst all the
resonators. A {\it single\/} transducer suffices to experimentally
determine the only monopole GW amplitude $\hat g^{(00)}(s)$, of course,
but (\ref{6.13}) provides the system response if more than one sensor is
mounted on the antenna for whatever reasons.

\subsection{Quadrupole gravitational radiation sensing}

We now consider the more interesting case of quadrupole motion sensing.
We thus take

\begin{equation}
   \Omega = \omega_{n2}\ ,\qquad (l_0=2)	\label{6.14}
\end{equation}
where $n\/$ labels the chosen harmonic ---most likely the first
($n\/$\,=\,1) or the second ($n\/$\,=\,2) in a practical system. The
evaluation of the $\Lambda$-matrix is now considerably more involved
\cite{serrano}, yet a remarkably elegant form is found for it:

\begin{equation}
  \hat\Lambda_a^{(lm)}(s;\omega_{n2}) = (-1)^N\,\sqrt{\frac{4\pi}{5}}\,
  a_{n2}\,\sum_{b=1}^J\,\left\{\sum_{\zeta_c\neq 0}\,\frac{1}{2}\left[
  \left(s^2+\omega_{c+}^2\right)^{-1} - \left(s^2+\omega_{c-}^2\right)^{-1}
  \right]\,\frac{v_a^{(c)}v_b^{(c)*}}{\zeta_c}\right\}\,
  Y_{2m}({\bf n}_b)\,\delta_{l2}    \label{6.15}
\end{equation}
where $v_a^{(c)}$ is the $c\/$-th normalised eigenvector of
$P_2({\bf n}_a\!\cdot\!{\bf n}_b)$, associated to the {\it non-null\/}
eigenvalue $\zeta_c^2$. Let us stress that equation (\ref{6.15}) explicitly
shows that at most 5 pairs of modes, of frequencies $\omega_{c\pm}$, couple
strongly to quadrupole GW amplitudes, {\it no matter how many resonators in
excess of 5 are mounted on the sphere\/}. The tidal overlap coefficients
$a_{2n}\/$ can also be calculated, and give for the first two harmonics
\cite{ls}

\begin{equation}
  \frac{a_{12}}{R} = 0.328\ ,\qquad\frac{a_{22}}{R} = 0.106   \label{6.16}
\end{equation}

The system response is thus

\begin{eqnarray}
  \hat q_a(s) & = & \eta^{-1/2}\,(-1)^J\,\sqrt{\frac{4\pi}{5}}\,a_{n2}\,
  \sum_{b=1}^J\,\left\{\sum_{\zeta_c\neq 0}\,\frac{1}{2}\left[
  \left(s^2+\omega_{c+}^2\right)^{-1} - \left(s^2+\omega_{c-}^2\right)^{-1}
  \right]\,\frac{v_a^{(c)}v_b^{(c)*}}{\zeta_c}\right\}\times
  \nonumber \\[0.5 em]  & & \hspace*{4 cm}
  \times\sum_{m=-2}^2\,Y_{2m}({\bf n}_b)\,\hat g^{(2m)}(s) + O(0)\ ,
  \qquad a=1\,\ldots,J	\label{6.17}
\end{eqnarray}

Equation (\ref{6.17}) is {\it completely general\/}, i.e., it is valid
for any resonator configuration over the sphere's surface, and for any
number of resonators. It describes precisely how all 5 GW amplitudes
$\hat g^{(2m)}(s)$ interact with all 5 strongly coupled system modes;
like before, {\it only quadrupole wave amplitudes\/} are seen in the
detector (to leading order) when $\Omega$\,=\,$\omega_{n2}$, even if
the incoming wave carries significant monopole energy at the frequencies
$\omega_{c\pm}$.

The degree of generality and algebraic simplicity of (\ref{6.17}) is new in
the literature. As we shall now see, it makes possible a systematic search
for different resonator distributions and their properties.

\section{The {\sl PHC\/} configuration}
\label{sec:PHC}

Merkowitz and Johnson's {\sl TIGA\/} \cite{jm93} is highly symmetric, and is
the minimal set with maximum degeneracy, i.e., all the non-null eigenvalues
$\zeta_a\/$ are equal. To accomplish this, however, 6 rather than 5
resonators are required on the sphere's surface. Since there are just
5 quadrupole GW amplitudes one may wonder whether there are alternative
layouts with {\it only\/} 5 resonators. Equation (\ref{6.17}) is completely
general, so it can be searched for an answer to this question. In reference
\cite{ls} we made a specific proposal, which we now describe in more detail.

In pursuing a search for 5 resonator sets we found that distributions having
a sphere diameter as an axis of {\it pentagonal symmetry\/}\footnote{
By this we mean resonators are placed along a {\it parallel\/} of the
sphere every 72$^\circ$.}
exhibit a rather appealing structure. More specifically, let the resonators
be located at the spherical positions

\begin{equation}
   \theta_a  = \alpha \qquad ({\rm all}\,\ a)\ ,\qquad
   \varphi_a = (a-1)\,\frac{2\pi}{5}\ ,\qquad a=1,\ldots,5
\end{equation}

The eigenvalues and eigenvectors of $P_2({\bf n}_a\!\cdot\!{\bf n}_b)$ are
easily calculated:

\begin{eqnarray}
   & \zeta_0^2 = \frac{5}{4}\,\left(3\,\cos^2\alpha-1\right)^2\ ,\qquad
       \zeta_1^2 = \zeta_{-1}^2 = \frac{15}{2}\,\sin^2\alpha\,\cos^2\alpha
       \ ,\qquad\zeta_2^2 = \zeta_{-2}^2 = \frac{15}{8}\,\sin^4\alpha  &
	\label{6.24.a} \\[1 em]
   & v_a^{(m)} = \sqrt{\frac{4\pi}{5}}\,\zeta_m^{-1}\,Y_{2m}({\bf n}_a)\ ,
       \qquad m=-2,\ldots,2\ ,\ \ a=1,\ldots,5	&
	\label{6.24.b}
\end{eqnarray}
so the $\Lambda$-matrix is also considerably simple in structure in this
case:

\begin{equation}
  \hat\Lambda_a^{(lm)}(s;\omega_{n2}) = -\sqrt{\frac{4\pi}{5}}\,a_{n2}\,
  \zeta_m^{-1}\,\frac{1}{2}\left[\left(s^2+\omega_{m+}^2\right)^{-1} -
  \left(s^2+\omega_{m-}^2\right)^{-1}\right]\,Y_{2m}({\bf n}_a)\,\delta_{l2}
  \qquad ({\sl PHC})     \label{6.25}
\end{equation}
where we have used the notation

\begin{equation}
  \omega_{m\pm}^2 = \Omega^2\,\left(1\pm\sqrt{\frac{5}{4\pi}}\,
  \left|A_{n2}(R)\right|\,\zeta_m\,\eta^{1/2}\right) + O(\eta)\ ,
  \qquad m=-2,\ldots,2
  \label{6.26}
\end{equation}

As we see from these formulas, the {\it five\/} expected pairs of
frequencies actually reduce to {\it three\/}, so pentagonal distributions
keep a certain degree of degeneracy, too. The most important distinguishing
characteristic of the general {\it pentagonal\/} layout is best displayed
by the explicit system response:

\begin{eqnarray}
  & \hat q_a(s) = -\eta^{-1/2}\,\sqrt\frac{4\pi}{5}\,a_{n2} & \left\{\,
  \frac{1}{2\zeta_0}\left[
  \left(s^2+\omega_{0+}^2\right)^{-1} - \left(s^2+\omega_{0-}^2\right)^{-1}
  \right]\,Y_{20}({\bf n}_a)\,\hat g^{(20)}(s)\right. \nonumber \\
  & & + \;\frac{1}{2\zeta_1}\left[
  \left(s^2+\omega_{1+}^2\right)^{-1} - \left(s^2+\omega_{1-}^2\right)^{-1}
  \right]\,\left[
     Y_{21}({\bf n}_a)\,\hat g^{(11)}(s) +
     Y_{2-1}({\bf n}_a)\,\hat g^{(1\,-1)}(s)\right] \label{6.27}  \\
  & & + \left.\frac{1}{2\zeta_2}\left[
  \left(s^2+\omega_{2+}^2\right)^{-1} - \left(s^2+\omega_{2-}^2\right)^{-1}
  \right]\,\left[
     Y_{22}({\bf n}_a)\,\hat g^{(22)}(s) +
     Y_{2-2}({\bf n}_a)\,\hat g^{(2\,-2)}(s)\right]\right\}
  \nonumber
\end{eqnarray}

This equation indicates that {\it different wave amplitudes selectively
couple to different detector frequencies\/}. This should be considered a
very remarkable fact, for it thence follows that simple inspection of the
system readout {\it spectrum\/}\footnote{
In a noiseless system, of course}
immediately reveals whether a given wave amplitude $\hat g^{2m}(s)$ is
present in the incoming signal or not.

Pentagonal configurations also admit {\it mode channels\/}, which are
easily constructed from (\ref{6.27}) thanks to the orthonormality property
of the eigenvectors (\ref{6.24.b}):

\begin{equation}
  \hat y^{(m)}(s)\equiv\sum_{a=1}^5\,v_a^{(m)*}\hat q_a(s) =
  \eta^{-1/2}\,a_{n2}\,
  \frac{1}{2}\left[\left(s^2+\omega_{m+}^2\right)^{-1} -
  \left(s^2+\omega_{m-}^2\right)^{-1}\right]\,\hat g^{(2m)}(s) + O(0)
  \ ,\qquad m=-2,\ldots,2
  \label{6.28}
\end{equation}

These are almost identical to the {\sl TIGA\/} mode channels \cite{jm95},
the only difference being that each mode channel comes now at a {\it single
specific\/} frequency pair $\omega_{m\pm}$.

{\it Mode channels\/} are fundamental in signal deconvolution algorithms
in noisy systems \cite{m98,lms}. Pentagonal resonator configurations should
thus be considered non-trivial candidates for a real GW detector.

\begin{figure}
\psfig{file=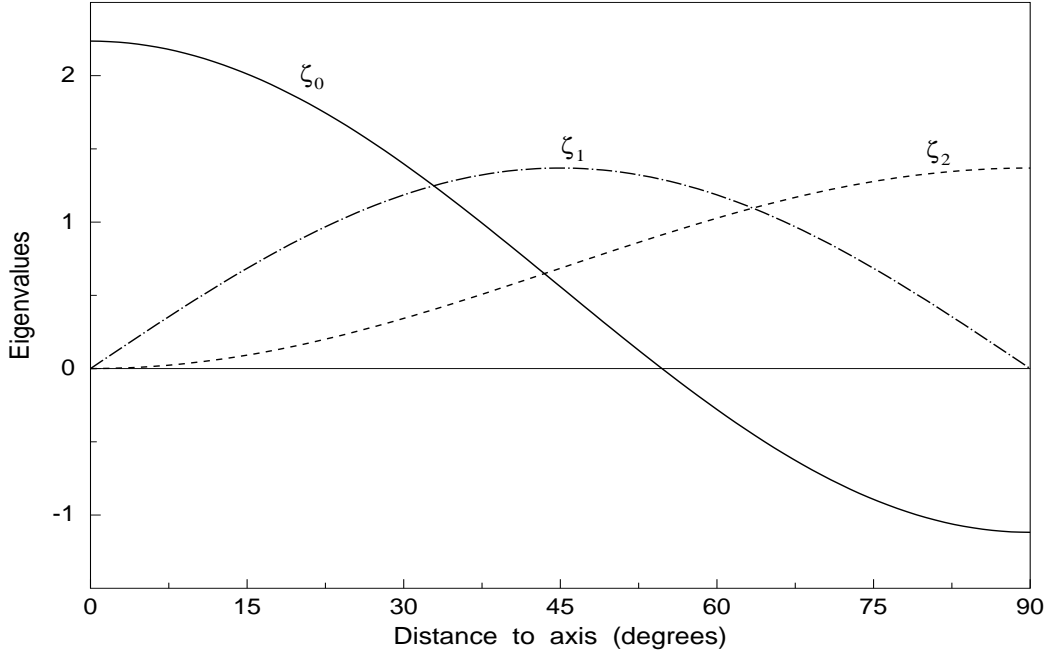,height=15cm,width=12cm,rheight=9.2cm,bbllx=-1.8cm,bblly=-3.8cm,bburx=18.4cm,bbury=24.7cm}
\caption{The three distinct eigenvalues $\zeta_m\/$ ($m\/$\,=\,0,1,2) as
functions of the distance of the resonator parallel's co-latitude $\alpha\/$
relative to the axis of symmetry of the distribution, cf. equation
(\protect\ref{6.24.a}).
\label{fig3}}
\end{figure}
\begin{figure}
\psfig{file=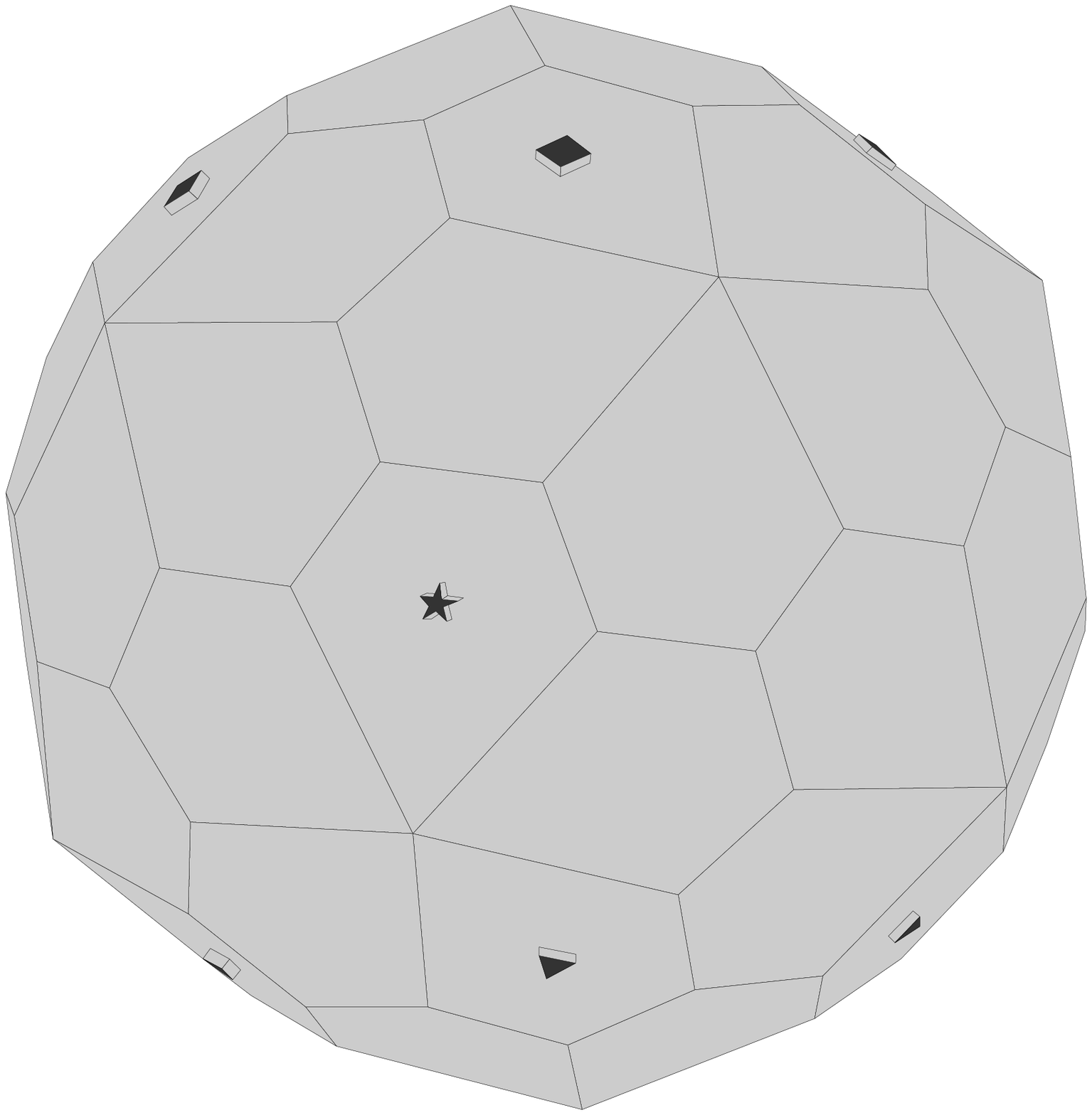,height=11cm,width=7cm}
\vspace*{-11 cm}
\hspace*{9.7 cm}
\psfig{file=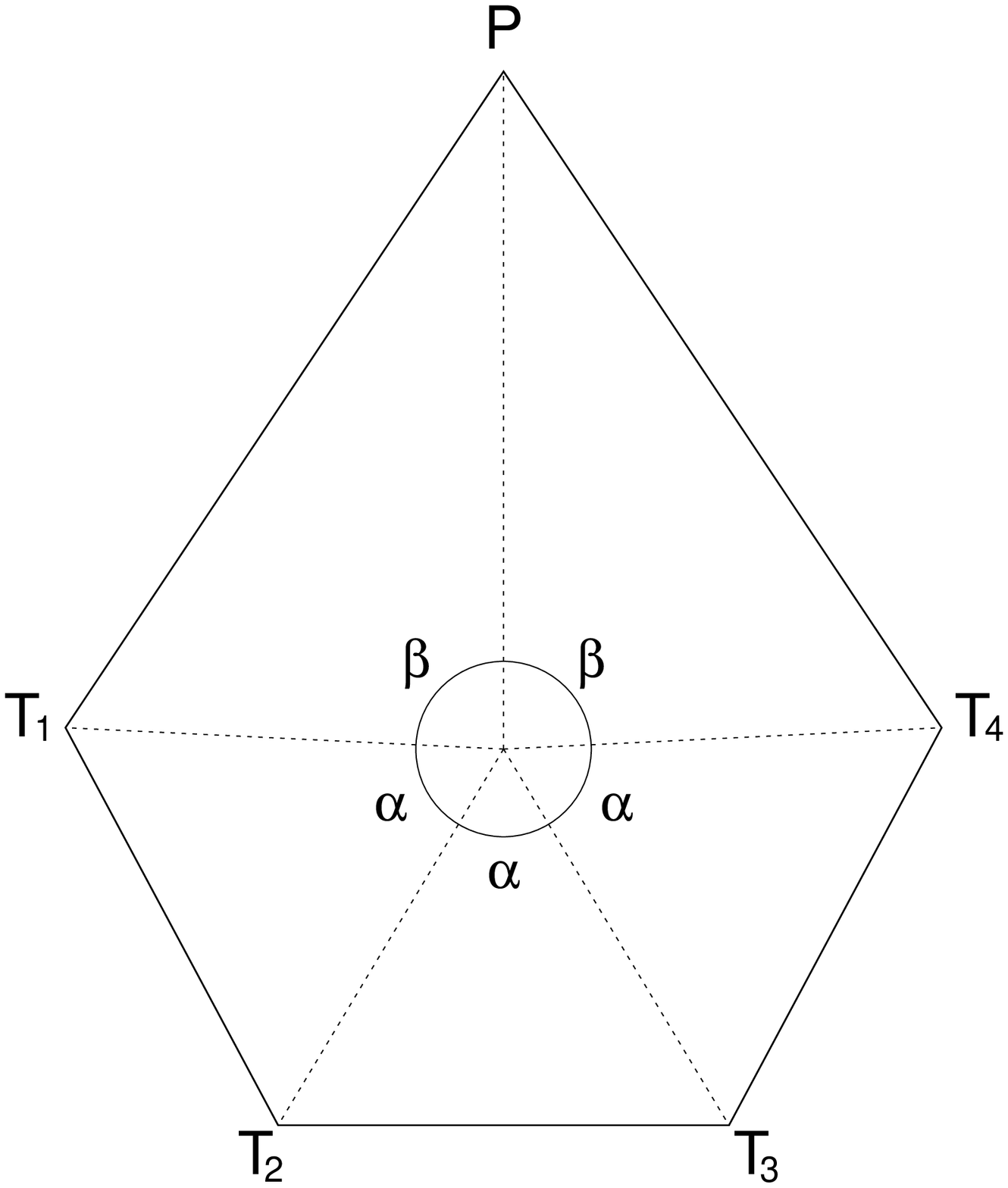,height=11cm,width=7cm}
\vspace*{-1.2 cm}
\caption{To the left, the {\it pentagonal hexacontahedron\/} shape. Certain
faces are marked to indicate resonator positions in a specific proposal
---see text--- as follows: a {\it square\/} for resonators tuned to the
first quadrupole frequency, a {\it triangle\/} for the second, and a
{\it star\/} for the monopole. On the right we see the (pentagonal) face
of the polyhedron. A few details about it: the confluence point of the
dotted lines at the centre is the tangency point of the {\it inscribed\/}
sphere to the {\sl PHC\/}; the labeled angles have values
$\alpha\/$\,=\,61.863$^\circ$, $\beta\/$\,=\,87.205$^\circ$; the angles at
the $T\/$-vertices are all equal, and their value is 118.1366$^\circ$,
while the angle at $P\/$ is 67.4536$^\circ$; the ratio of a long edge
(e.g. $PT_1$) to a short one (e.g. $T_1T_2$) is 1.74985, and the radius of
the inscribed sphere is {\it twice\/} the long edge of the pentagon,
$R\/$\,=\,2\,$PT_1$.	\label{fig4}}
\end{figure}
\begin{figure}
\psfig{file=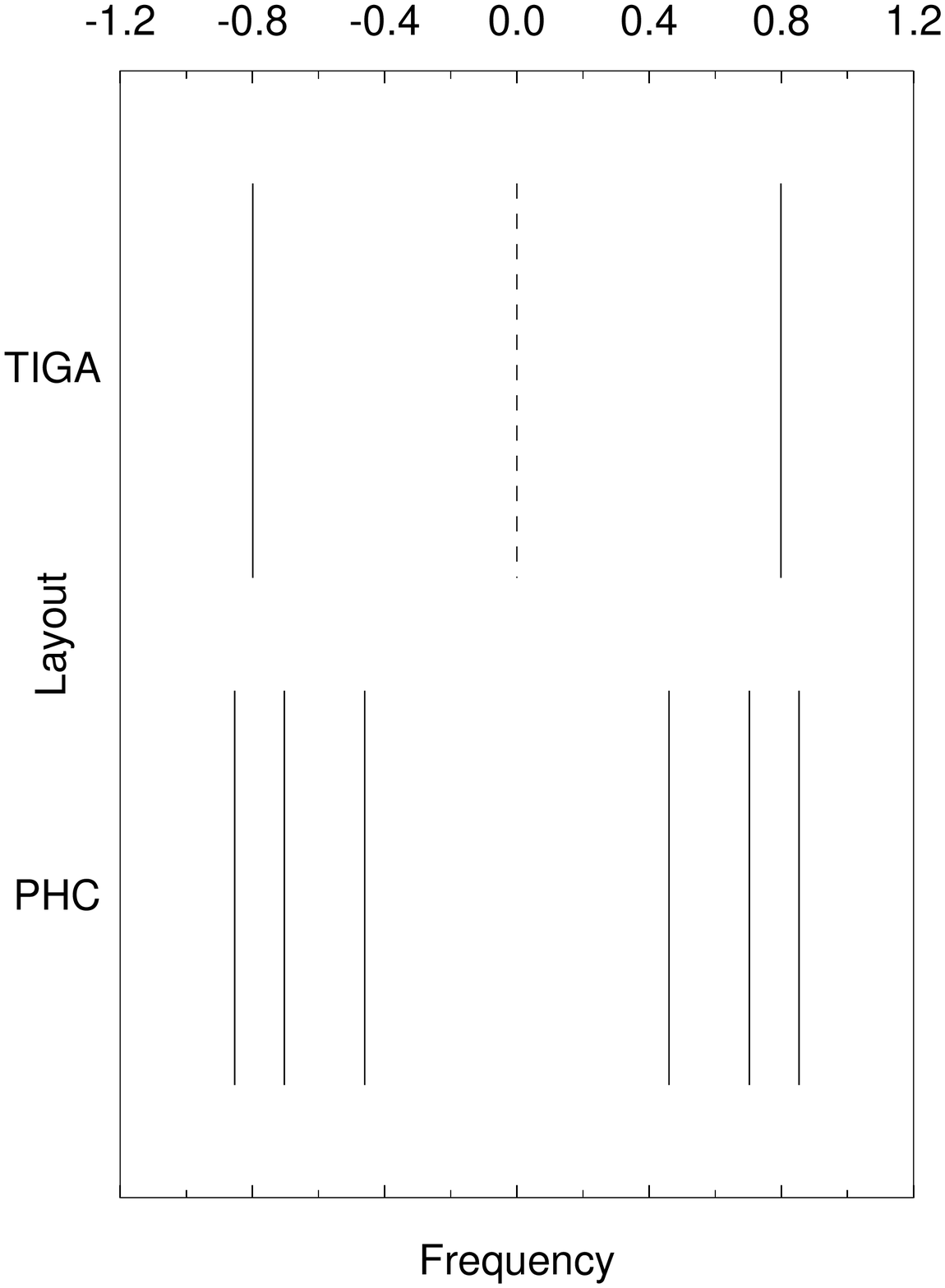,height=12cm,width=15cm,rheight=9.7cm,bbllx=-1cm,bblly=-5cm,bburx=20cm,bbury=25.8cm}
\caption{Compared line spectrum of a coupled {\sl TIGA\/} and a {\sl PHC\/}
resonator layout in an ideally spherical system. The weakly coupled central
frequency in the {\sl TIGA\/} is drawn dashed. The frequency pair is 5-fold
degenerate for this layout, while the two outer pairs of the {\sl PHC\/}
are doubly degenerate each, and the inner pair is non-degenerate. Units in
abscissas are $\eta^{1/2}\Omega$, and the central value, labeled 0.0,
corresponds to $\Omega$.
\label{fig5}}
\end{figure}

Based on these facts one may next ask which is a suitable transducer
distribution with an axis of pentagonal symmetry. In Figure \ref{fig3} we
give a plot of the eigenvalues (\ref{6.24.a}) as a function of $\alpha\/$,
the angular distance of the resonator set from the symmetry axis. Several
criteria may be adopted to select a specific choice in view of this graph.
An interesting one was proposed by us in reference \cite{ls} with the
following argument. If for ease of mounting, stability, etc., it is
desirable to have the detector milled into a close-to-spherical
{\it polyhedric\/} shape\footnote{
This is the philosophy suggested and experimentally implemented by
Merkowitz and Johnson at {\sl LSU\/}.}
then polyhedra with axes of pentagonal symmetry must be searched. The
number of quasi regular {\it convex\/} polyhedra is of course finite
---there actually are only 18 of them \cite{pacoM,tsvi}---, and we found
a particularly appealing one in the so called {\it pentagonal
hexacontahedron\/} ({\sl PHC\/}), which we see in Figure \ref{fig4}, left.
This is a 60 face polyhedron, whose faces are the identical {\it irregular
pentagons\/} of Figure \ref{fig4}, right. The {\sl PHC\/} admits an
{\it inscribed sphere\/} which is tangent to each face at the central
point marked in the Figure. It is clearly to this point that a resonator
should be attached so as to simulate an as perfect as possible spherical
distribution.

The {\sl PHC\/} is considerably spherical: the ratio of its volume to that
of the inscribed sphere is 1.057, which quite favourably compares to the
value of 1.153 for the ratio of the circumscribed sphere to the TI volume.
If we now request that the frequency pairs $\omega_{m\pm}$ be as
{\it evenly spaced\/} as possible, compatible with the {\sl PHC\/} face
orientations, then we must choose $\alpha\/$\,=\,67.617$^\circ$, whence

\begin{equation}
 \omega_{0\pm} = \omega_{12}\,\left(1\pm 0.5756\,\eta^{1/2}\right)\ ,\ \ \ 
 \omega_{1\pm} = \omega_{12}\,\left(1\pm 0.8787\,\eta^{1/2}\right)\ ,\ \ \ 
 \omega_{2\pm} = \omega_{12}\,\left(1\pm 1.0668\,\eta^{1/2}\right)
 \label{6.29}
\end{equation}
for instance for $\Omega$\,=\,$\omega_{12}$, the first quadrupole harmonic.
In Figure \ref{fig5} we display this frequency spectrum together with the
multiply degenerate {\it TIGA\/} for comparison.

The criterion leading to the {\sl PHC\/} proposal is of course not unique,
and alternatives can be considered. For example, if the 5 faces of a
regular icosahedron are selected for sensor mounting
($\alpha\/$\,=\,63.45$^\circ$) then a four-fold degenerate pair plus a
single non-degenerate pair is obtained; if the resonator parallel is
50$^\circ$ or 22.6$^\circ$ away from the ``north pole'' then the three
frequencies $\omega_{0+}$, $\omega_{1+}$, and $\omega_{2+}$ are equally
spaced; etc. The number of choices is virtually infinite if the sphere is
not milled into a polyhedric shape \cite{ts,grg}.

Let us finally recall that the complete {\sl PHC\/} proposal \cite{ls} was
made with the idea of building an as complete as possible spherical GW
antenna, which amounts to making it sensitive at the first {\it two\/}
quadrupole frequencies {\it and\/} at the first monopole one. This would
take advantage of the good sphere cross section at the second quadrupole
harmonic \cite{clo}, and would enable measuring (or thresholding)
eventual monopole GW radiation. Now, the system {\it pattern matrix\/}
$\hat\Lambda_a^{(lm)}(s;\Omega)$ has {\it identical structure\/} for all the
harmonics of a given $l\/$ series ---see (\ref{6.12}) and (\ref{6.15})---,
and so too identical criteria for resonator layout design apply to either
set of transducers, respectively tuned to $\omega_{12}$ and $\omega_{22}$.
The {\sl PHC\/} proposal is best described graphically in Figure \ref{fig4},
left: a {\it second\/} set of resonators, tuned to the second quadrupole
harmonic $\omega_{22}$ can be placed in an equivalent position in the
``southern hemisphere'', and an eleventh resonator tuned to the first
monopole frequency $\omega_{10}$ is added at an arbitrary position. It is not
difficult to see, by the general methods outlined earlier on in this paper,
that cross interaction between these three sets of resonators is only
{\it second order\/} in $\eta^{1/2}\/$, therefore weak.

A spherical GW detector with such a set of altogether 11 transducers would
be a very complete multi-mode multi-frequency device with an unprecedented
capacity as an individual antenna. Amongst other it would practically enable
monitoring of coalescing binary {\it chirp\/} signals by means of a rather
robust double passage method \cite{cf}, a prospect which was considered
so far possible only with broadband long baseline laser interferometers
\cite{klm1,klm2}, and is almost unthinkable with currently operating
cylindrical bars.

\section{A calibration signal: hammer stroke}
\label{sec:hs}

This section is a brief digression from the main streamline of the paper.
We propose to assess now the system response to a particular, but useful,
calibration signal: a perpendicular {\it hammer stroke\/}.

We first go back to equation (\ref{3.16}) and replace
$\hat u_a^{\rm external}(s)$ in its rhs with that corresponding to a
hammer stroke, which is easily calculated ---cf.\ appendix~\ref{app:a}:

\begin{equation}
  \hat u_a^{\rm stroke}(s) = -\sum_{nl}\,\frac{f_0}{s^2+\omega_{nl}^2}\,
  \left|A_{nl}(R)\right|^2\,P_l({\bf n}_a\!\cdot\!{\bf n}_0)\ ,\qquad
  a=1,\ldots,J  \label{7.1}
\end{equation}
where ${\bf n}_0$ are the spherical coordinates of the hit point on the
sphere, and $f_0$\,$\equiv$\,${\bf n}_0\!\cdot\!{\bf f}_0/{\cal M\/}$.
Clearly, the hammer stroke excites {\it all\/} of the sphere's vibration
eigenmodes, as it has a completly flat spectrum.

The coupled system resonances are again those calculated in
appendix~\ref{app:b}. The same procedures described in section~\ref{sec:srgw}
for a GW excitation can now be pursued to obtain

\begin{eqnarray}
  \hat q_a(s) & = & \eta^{-1/2}\,(-1)^{J-1}\,\sqrt{\frac{2l+1}{4\pi}}
  \,f_0\,\left|A_{nl}(R)\right|\,\times  \nonumber \\
  & \times & \sum_{b=1}^J\,\left\{\sum_{\zeta_c\neq 0}\,\frac{1}{2}\left[
  \left(s^2+\omega_{c+}^2\right)^{-1}-\left(s^2+\omega_{c-}^2\right)^{-1}
  \right]\,\frac{v_a^{(c)}v_b^{(c)*}}{\zeta_c}\right\}\,
  P_l({\bf n}_b\!\cdot\!{\bf n}_0) + O(0)\ ,
  \qquad a=1\,\ldots,J	\label{7.2}
\end{eqnarray}
when the system is tuned to the $nl\/$-th spheroidal harmonic, i.e.,
$\Omega$\,=\,$\omega_{nl}$. It is immediately seen from here that the
system response to this signal when the resonators are tuned to a
{\it monopole\/} frequency is given by

\begin{equation}
  \hat q_a(s) = \eta^{-1/2}\,(-1)^{J-1}\,\frac{f_0}{\sqrt{4\pi J}}\,
  \left|A_{n0}(R)\right|\,\frac{1}{2}\left[\left(
  s^2+\omega_+^2\right)^{-1}-\left(s^2+\omega_-^2\right)^{-1}\right]
  \ ,\qquad \Omega=\omega_{n0}
  \label{7.3}
\end{equation}
an expression which holds for all $a\/$, and is independent of either the
resonator layout or the hit point, which in particular prevents any
determination of the latter, as obviously expected. The frequencies
$\omega_\pm$ are those of (\ref{6.11}), and we find here again a global
factor $J^{-1/2}$, as also expected.

We consider next the situation when quadrupole tuning is implemented,
$\Omega$\,=\,$\omega_{n2}$. We shall however do so only for the {\sl PHC\/}
and {\sl TIGA\/} configurations, as more general considerations are not
quite as interesting at this point.

\begin{figure}
\psfig{file=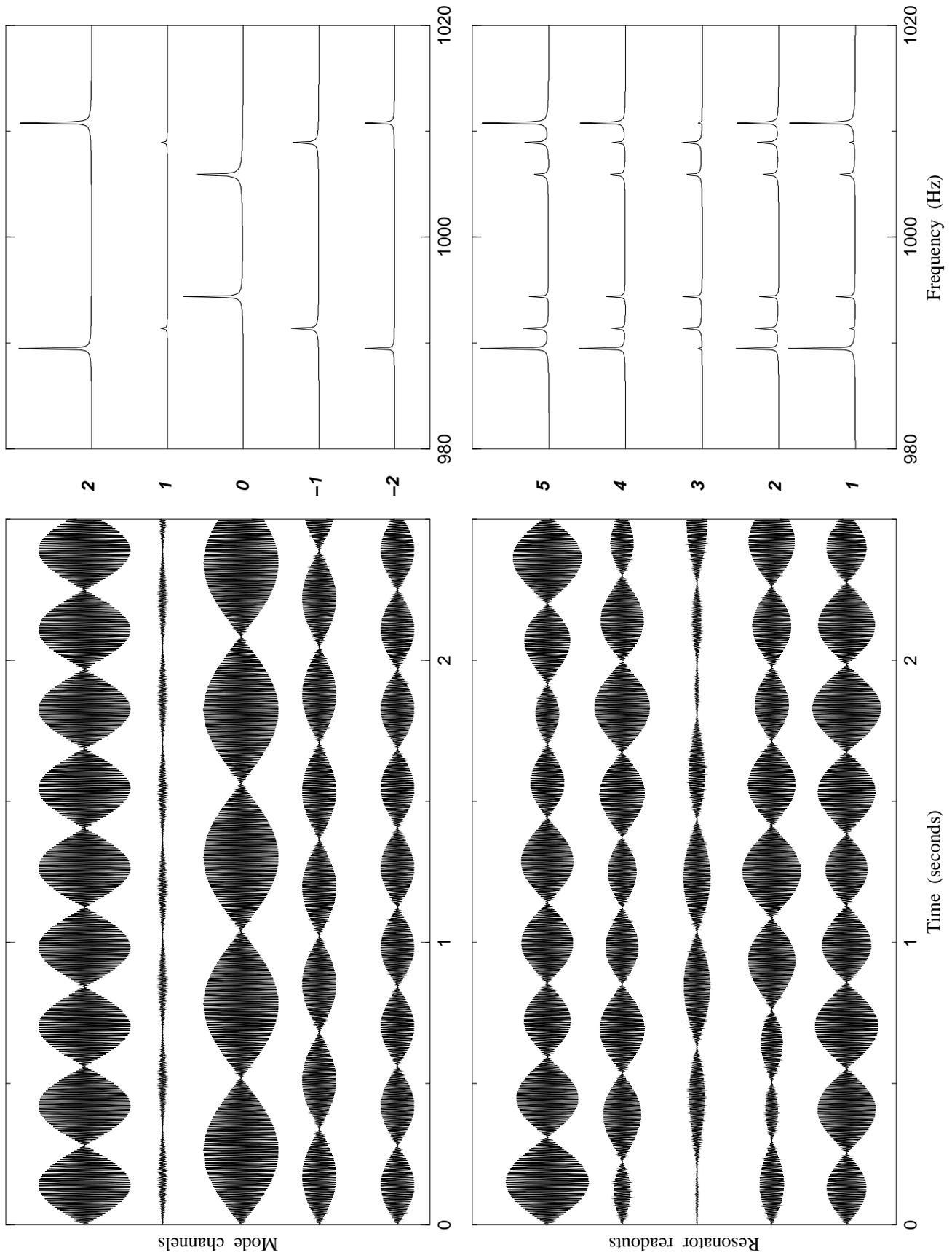,height=22cm,width=15cm,rheight=23cm,bbllx=1.1cm,bblly=3.3cm,bburx=17.5cm,bbury=27cm}
\caption{Simulated response of a {\sl PHC\/} to a hammer stroke: the time
series and their respective spectra, both for direct resonator readouts and
mode channels. Note that while the former are {\it not\/} simple beats, the
latter are.	\label{fig7}}
\end{figure}

\subsection{{\sl PHC\/} and {\sl TIGA\/} response to a hammer stroke}

Expanding equation \ref{7.2} by substitution of the eigenvalues $\zeta_m\/$
and eigenvectors $v^{(m)}_a\/$ of the {\sl PHC\/}, one readily finds that
the system response is given by

\begin{equation}
  \hat q_a(s) = \eta^{-1/2}\,f_0\,\sqrt{\frac{4\pi}{5}}\,
  \left|A_{n2}(R)\right|\,\sum_{m=-2}^2\,\frac{1}{2}\left[\left(
  s^2+\omega_{m+}^2\right)^{-1}-\left(s^2+\omega_{m-}^2\right)^{-1}\right]
  \,\zeta_m^{-1}\,Y_{2m}({\bf n}_a)\,Y_{2m}^*({\bf n}_0)
  \ ,\qquad a=1,\ldots,5
  \label{7.7}
\end{equation}
and the mode channels by

\begin{equation}
  \hat y^{(m)}(s) = \eta^{-1/2}\,f_0\,\left|A_{n2}(R)\right|
  \frac{1}{2}\left[\left(s^2+\omega_{m+}^2\right)^{-1} -
  \left(s^2+\omega_{m-}^2\right)^{-1}\right]\,Y_{2m}^*({\bf n}_0)
  \ ,\ \ m=-2,\ldots,2	\label{7.8}
\end{equation}

These equations indicate that the system response $q_a(t)$ is a
{\it superposition of three different beats\/}\footnote{
A {\it beat\/} is a modulated oscillation of the form
$\sin\frac{1}{2}(\omega_+-\omega_-)t\,\cos\Omega t$, where $\omega_+$
and $\omega_-$ are nearby frequencies, and
$\omega_+$\,$+$\,$\omega_-$\,=\,2\,$\Omega$. The Laplace transform of such
function of time is precisely $(\Omega/2)$\,$\left[\left(
s^2+\omega_+^2\right)^{-1}-\left(s^2+\omega_-^2\right)^{-1}\right]$, up to
higher order terms in the difference $\omega_+$\,$-$\,$\omega_-$, which in
our case is proportional to $\eta^{1/2}$.},
while the mode channels are {\it single\/} beats each, but with {\it differing
modulation frequencies\/}. This is represented graphically in Figure
\ref{fig7}, where we see the result of a numerical simulation of the
{\sl PHC\/} response to a hammer stroke, delivered to the solid at a given
location. The readouts $q_a(t)$ are somewhat complex time series, whose
frequency spectrum shows {\it three pairs of peaks\/} ---in fact, the
{\it lines\/} in the ideal spectrum of Figure \ref{fig5}. The mode channels
on the other hand are {\it pure beats\/}, whose spectra consist of the
{\it individually separate\/} pairs of the just mentioned peaks.

The response of the {\sl TIGA\/} layout to a hammer stroke has been described
in detail by Merkowitz and Johnson ---see e.g.\ reference \cite{jm97}. Our
formalism does of course recover the results obtained by those authors; in
the notation of this paper, we have

\begin{eqnarray}
  \hat q_a(s) & = & -\eta^{-1/2}\,\frac{5}{\sqrt{24\pi}}\,
  f_0\,\left|A_{n2}(R)\right|\,\frac{1}{2}\left[
  \left(s^2+\omega_+^2\right)^{-1}-\left(s^2+\omega_-^2\right)^{-1}\right]
  \,P_2({\bf n}_a\!\cdot\!{\bf n}_0)\ ,\qquad {\sl TIGA}
  \label{7.4.a} \\
  \hat y^{(m)}(s) & = & -\eta^{-1/2}\,f_0\,\left|A_{n2}(R)\right|
  \frac{1}{2}\left[\left(s^2+\omega_+^2\right)^{-1} -
  \left(s^2+\omega_-^2\right)^{-1}\right]\,Y_{2m}^*({\bf n}_0)
  \ ,\ \  m=-2,\ldots,2	\label{7.5}
\end{eqnarray}
for the system response and the mode channels, respectively, where

\begin{equation}
  \omega_\pm^2 = \omega_{n2}^2\,\left(1\pm\sqrt{\frac{3}{2\pi}}\,
  \left|A_{n2}(R)\right|\eta^{1/2}\right) + O(\eta)\ ,
  \qquad a=1,\ldots,6
  \label{6.19}
\end{equation}
are the five-fold degenerate frequency pairs corresponding to the {\sl TIGA\/}
distribution. Comparison of the mode channels shows that they are identical
for {\sl PHC\/} and {\sl TIGA\/}, except that the former come at different
frequencies depending on the index $m\/$. One might perhaps say that the
{\sl PHC\/} gives rise to a sort of ``Zeeman splitting'' of the {\sl TIGA\/}
degenerate frequencies, which can be attributed to an {\it axial symmetry
breaking\/} of that resonator distribution: the {\sl PHC\/} mode channels
partly split up the otherwise degenerate multiplet into its components.

\section{Symmetry defects}
\label{sec:symdef}

So far we have made the assumption that the sphere is perfectly symmetric,
that the resonators are identical, that their locations on the sphere's
surface are ideally accurate, etc. This is of course unrealistic. So we
propose to address now how such departures from ideality affect the
system behaviour. As we shall see, the system is rather {\it robust\/},
in a sense to be made precise shortly, against a number of small defects.

In order to {\it quantitatively\/} assess ideality failures we shall adopt
a philosophy which is naturally suggested by the results already obtained
in an ideal system. It is as follows.

As we have seen in previous sections, the solution to the general equations
(\ref{3.16}) must be given as a {\it perturbative\/} series expansion in
ascending powers of the small quantity $\eta^{1/2}$. This is clearly a fact
{\it not\/} related to the system's symmetries, so it will survive symmetry
breakings. It is therefore appropriate to {\it parametrize\/} deviations
from ideality in terms of suitable powers of $\eta^{1/2}$, in order to
address them {\it consistently with the order of accuracy of the series
solution to the equations of motion\/}. An example will better illustrate
the situation.

In a {\it perfectly ideal\/} spherical detector the system frequencies
are given by equations (\ref{5.2}). Now, if a small departure from e.g.
spherical symmetry is present in the system then we expect that a
correspondingly small correction to those equations will be required.
Which specific correction to the formula will actually happen can be
{\it qualitatively\/} assessed by a {\it consistency\/} argument: if
symmetry defects are of order $\eta^{1/2}$ then equations (\ref{5.2}) will
be significantly altered in their $\eta^{1/2}$ terms; if on the other hand
such defects are of order $\eta\/$ or smaller then any modifications to
equations (\ref{5.2}) will be swallowed into the $O(0)$ terms, and the
more important $\eta^{1/2}$ terms will remain unaffected by the symmetry
failure. We will say in the latter case that the system is {\it robust\/}
to that ideality breaking.

More generally, this argument can be extended to see that the only system
defects standing a chance to have any influences on lowest order ideal
system behaviour are defects of order $\eta^{1/2}$ relative to an ideal
configuration. Defects of such order are however {\it not necessarily
guaranteed\/} to be significant, and a specific analysis is required for
each specific parameter in order to see whether or not the system response
is {\it robust\/} against the considered parameter deviations.

We therefore proceed as follows. Let $P\/$ be one of the system parameters,
e.g. a sphere frequency, or a resonator mass or location, etc. Let
$P_{\rm ideal}$ be the {\it numerical value\/} this parameter has in an
ideal detector, and let $P_{\rm real}$ be its value in the real case. These
two will be assumed to differ by terms of order $\eta^{1/2}$, or

\begin{equation}
  P_{\rm real} = P_{\rm ideal}\,(1+p\,\eta^{1/2})     \label{8.1}
\end{equation}

For a given system, $p\/$ is readily determined adopting (\ref{8.1}) as the
{\it definition\/} of $P_{\rm real}$, once a suitable {\it hypothesis\/}
has been made as to which is the value of $P_{\rm ideal}$. In order for
the following procedure to make sensible sense it is clearly required that
$p\/$ be of order 1 or, at least, appreciably larger than $\eta^{1/2}$.
Should $p\/$ thus calculated from (\ref{8.1}) happen to be too small, i.e.,
of order $\eta^{1/2}$ itself or smaller, then the system will be considered
{\it robust\/} as regards the affected parameter.

We now apply this criterion to various departures from ideality.

\subsection{The suspended sphere  \label{s8.1}}

An earth based observatory obviously requires a {\it suspension
mechanism\/} for the large sphere. If a {\it nodal point\/} suspension
is e.g.\ selected then a diametral {\it bore\/} has to be drilled across
the sphere \cite{phd}. The most immediate consequence of this is that
spherical symmetry is broken, what in turn results in {\it degeneracy
lifting\/} of the free spectral frequencies $\omega_{nl}\/$, which now
{\it split\/} up into multiplets $\omega_{nlm}\/$
($m\/$\,=\,$-l\/$,...,$l\/$). The resonators' frequency $\Omega$
{\it cannot\/} therefore be matched to {\it the\/} frequency
$\omega_{n_0l_0}$, but at most to {\it one\/} of the
$\omega_{n_0l_0m}\/$'s. In this subsection we keep the hypothesis that
all the resonators are identical ---we shall relax it later---, and
assume that $\Omega$ falls {\it within\/} the span of the multiplet
of the $\omega_{n_0l_0m}\/$'s. Then we write

\begin{equation}
  \omega_{n_0l_0m}^2 = \Omega^2\,(1+p_m\,\eta^{1/2})\ ,\qquad
  m=-l_0,\ldots,l_0     \label{8.2}
\end{equation}

We now search for the coupled frequencies, i.e., the roots of equation
(\ref{3.18}). The kernel matrix $\hat K_{ab}(s)$ is however no longer
given by (\ref{4.2}), due the removed degeneracy of $\omega_{nl}\/$, and
we must stick to its general expression (\ref{3.17}), or

\begin{equation}
  \hat K_{ab}(s) = \sum_{nlm}\,\frac{\Omega_b^2}{s^2+\omega_{nlm}^2}\,
   \left|A_{nl}(R)\right|^2\,\frac{2l+1}{4\pi}\,
   Y_{lm}^*({\bf n}_a)\,Y_{lm}({\bf n}_b) \equiv
   \sum_{nlm}\,\frac{\Omega_b^2}{s^2+\omega_{nlm}^2}\,\chi_{ab}^{(nlm)}
   \label{8.3}
\end{equation}

Following the steps of appendix \ref{app:a} we now seek the roots of the
equation

\begin{equation}
  \det\,\left[\delta_{ab} + \eta\,\sum_{m=-l_0}^{l_0}\,
   \frac{\Omega^2s^2}{(s^2+\Omega^2)(s^2+\omega_{n_0l_0m}^2)}
   \,\chi_{ab}^{(n_0l_0m)} + \eta\,\sum_{nl\neq n_0l_0,m}\,
   \frac{\Omega^2s^2}{(s^2+\Omega^2)(s^2+\omega_{nlm}^2)}\,
   \chi_{ab}^{(nlm)}\right] = 0
  \label{8.4}
\end{equation}

Since $\Omega$ relates to $\omega_{n_0l_0m}\/$ through equation (\ref{8.2})
we see that the roots of (\ref{8.4}) fall again into either of the two
categories (\ref{4.11.a})-(\ref{4.11.b})(see Appendix~\ref{app:b}), i.e.,
roots close to $\pm i\Omega$ and roots close to $\pm i\omega_{nlm}\/$
($nl\/$\,$\neq$\,$n_0l_0$). We shall exclusively concentrate on the former
now. Direct substitution of the series (\ref{4.11.a}) into (\ref{8.4}) yields
the following equation for the coefficient $\chi_{\frac{1}{2}}$:

\begin{equation}
  \det\left[\delta_{ab} - \frac{1}{\chi_\frac{1}{2}}\,\sum_{m=-l_0}^{l_0}
  \,\frac{\chi_{ab}^{(n_0l_0m)}}{\chi_\frac{1}{2}-p_m}\right] = 0
  \label{8.5}
\end{equation}

This is a variation of (\ref{5.1}), to which it reduces when
$p_m\/$\,=\,0, i.e., when there is full degeneracy.

The solutions to (\ref{8.5}) no longer come in symmetric pairs, like
(\ref{5.2}). Rather, there are 2$l_0$+1+$J\/$ of them, with a
{\it maximum\/} number of 2(2$l_0$+1) non-identically zero roots if
$J\/$\,$\geq$\,2$l_0$+1\footnote{
This is a {\it mathematical fact\/}, whose proof is relatively cumbersome,
and will be omitted here; we just mention that it has its origin in the
linear dependence of more than 2$l_0$+1 spherical harmonics of order $l_0$.}.
For example, if we choose to select the resonators' frequency close to a
quadrupole multiplet ($l_0$\,=\,2) then (\ref{8.5}) has at most 5+$J\/$
non-null roots, {\it with a maximum ten\/} no matter how many resonators
in excess of 5 we attach to the sphere. Modes associated to null roots of
(\ref{8.5}) can be seen to be {\it weakly coupled\/}, just like in a free
sphere, i.e., their amplitudes are smaller than those of the strongly
coupled ones by factors of order $\eta^{1/2}$.

In order to assess the reliability of this method we have applied it to
see what are its predictions for a {\it real system\/}. To this end, data
taken with the {\sl TIGA\/} prototype at {\sl LSU\/}\footnote{
These data are contained in reference \protect\cite{phd}, and we want to
express our gratitude to Stephen Merkowitz for kindly handing them to us.}
were used to confront with. The {\sl TIGA\/} was drilled and suspended
from its centre, so its first quadrupole frequency split up into a
multiplet of five frequencies. Their reportedly measured values are

\begin{equation}
  \omega_{120} = 3249\ {\rm Hz}\ ,\ \ 
  \omega_{121} = 3238\ {\rm Hz}\ ,\ \ 
  \omega_{12\,-1} = 3236\ {\rm Hz}\ ,\ \ 
  \omega_{122} = 3224\ {\rm Hz}\ ,\ \ 
  \omega_{12\,-2} = 3223\ {\rm Hz}\ ,\ \ 
  \label{8.6}
\end{equation}

All 6 resonators were equal, and had the following characteristic
frequency and mass, respectively:

\begin{equation}
  \Omega = 3241\ {\rm Hz}\ ,\qquad\eta = \frac{1}{1762.45}
  \label{8.7}
\end{equation}

Substituting these values into (\ref{8.2}) it is seen that

\begin{equation}
  p_0=0.2075\ ,\ \   p_1=-0.0777\ ,\ \   p_{-1}=-0.1036\ ,\ \ 
  p_2=-0.4393\ ,\ \  p_{-2}=-0.4650
  \label{8.8}
\end{equation}

\begin{table}
\label{t1}
\caption{Numerical values of measured and theoretically predicted
frequencies (in Hz) for the {\sl TIGA\/} prototype with varying number of
resonators. Percent differences are also shown. The {\it calculated\/}
values of the tuning and free multiplet frequencies are taken {\it by
definition\/} equal to the measured ones, and quoted in brackets. In
square brackets the frequency of the {\it weakly coupled\/} sixth mode
in the full, 6~resonator {\sl TIGA\/} layout. These data are plotted in
Figure \protect\ref{fig8}.}
\begin{center}
\begin{tabular}{lccc||lccc}
Item & Measured & Calculated  & \% difference &
Item & Measured & Calculated  & \% difference \\
\hline Tuning & 3241 & (3241) & (0.00) &
4 resonators & 3159 & 3155 & $-0.12$ \\
Free multiplet & 3223 & (3223)  & (0.00) &
             & 3160 & 3156 & $-0.11$ \\
               & 3224 & (3224)  & (0.00) &
             & 3168 & 3165 & $-0.12$ \\
               & 3236 & (3236)  & (0.00) &
             & 3199 & 3198 & $-0.05$ \\
               & 3238 & (3238)  & (0.00) &
             & 3236 & 3236 & $ 0.00$ \\
               & 3249 & (3249)  & (0.00) &
             & 3285 & 3286 & $ 0.03$ \\
1 resonator  & 3167 & 3164 & $-0.08$ &
             & 3310 & 3310 & $ 0.00$ \\
             & 3223 & 3223 & $ 0.00$ &
             & 3311 & 3311 & $ 0.00$ \\
             & 3236 & 3235 & $-0.02$ &
             & 3319 & 3319 & $ 0.00$ \\
             & 3238 & 3237 & $-0.02$ &
5 resonators & 3152 & 3154 & $ 0.08$ \\
             & 3245 & 3245 & $ 0.00$ &
             & 3160 & 3156 & $-0.14$ \\
             & 3305 & 3307 & $ 0.06$ &
             & 3163 & 3162 & $-0.03$ \\
2 resonators & 3160 & 3156 & $-0.13$ &
             & 3169 & 3167 & $-0.08$ \\
             & 3177 & 3175 & $-0.07$ &
             & 3209 & 3208 & $-0.02$ \\
             & 3233 & 3233 & $ 0.00$ &
             & 3268 & 3271 & $ 0.10$ \\
             & 3236 & 3236 & $ 0.00$ &
             & 3304 & 3310 & $ 0.17$ \\
             & 3240 & 3240 & $ 0.00$ &
             & 3310 & 3311 & $ 0.03$ \\
             & 3302 & 3303 & $ 0.03$ &
             & 3313 & 3316 & $ 0.10$ \\
             & 3311 & 3311 & $ 0.00$ &
             & 3319 & 3321 & $ 0.06$ \\
3 resonators & 3160 & 3155 & $-0.15$ &
6 resonators & 3151 & 3154 & $ 0.11$ \\
             & 3160 & 3156 & $-0.13$ &
             & 3156 & 3155 & $-0.03$ \\
             & 3191 & 3190 & $-0.02$ &
             & 3162 & 3162 & $ 0.00$ \\
             & 3236 & 3235 & $-0.02$ &
             & 3167 & 3162 & $-0.14$ \\
             & 3236 & 3236 & $ 0.00$ &
             & 3170 & 3168 & $-0.07$ \\
             & 3297 & 3299 & $ 0.08$ &
             & [3239] & [3241] & [0.06] \\
             & 3310 & 3311 & $ 0.02$ &
             & 3302 & 3309 & $ 0.23$ \\
             & 3311 & 3311 & $ 0.00$ &
             & 3308 & 3310 & $ 0.06$ \\
             &  &  &  &
             & 3312 & 3316 & $ 0.12$ \\
             &  &  &  &
             & 3316 & 3317 & $ 0.02$ \\
             &  &  &  &
             & 3319 & 3322 & $ 0.10$
\end{tabular}
\end{center}
\end{table}

\begin{figure}
\psfig{file=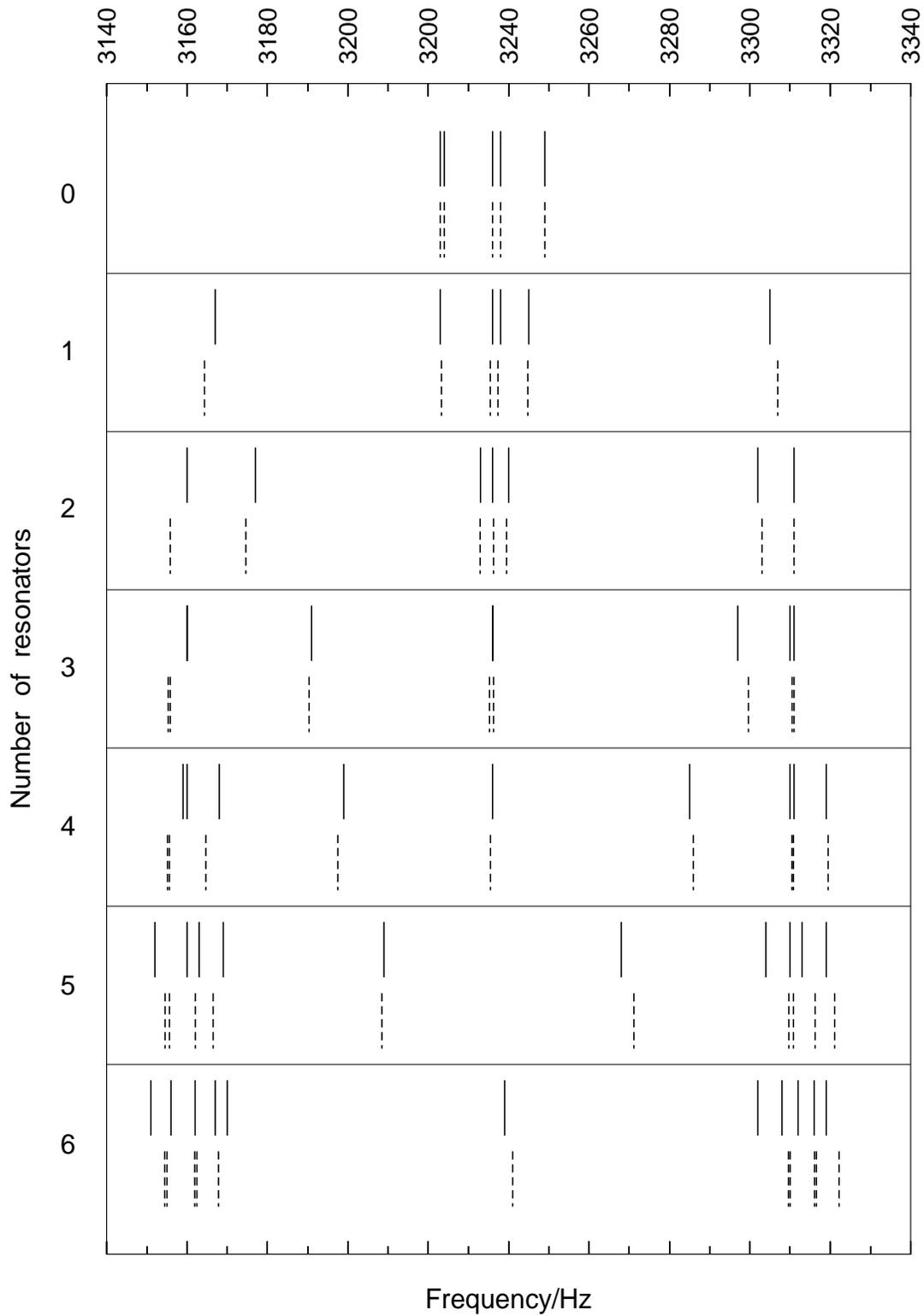,height=22cm,width=15cm,rheight=22cm}
\caption{The frequency spectrum of the {\sl TIGA\/} distribution as
resonators are progressively added from none to 6. Continuous lines
correspond to measured values, and dashed lines correspond to their
$\eta^{1/2}$ theoretical estimates with equation~(\protect\ref{8.5}).
\label{fig8}}
\end{figure}

Equation (\ref{8.5}) can now be readily solved once the resonator
positions are fed into the matrices $\chi_{ab}^{(12m)}$. Such positions
correspond to the pentagonal faces of a truncated icosahedron. Merkowitz
\cite{phd} gives a complete account of all the measured system frequencies
as resonators are progressively attached to the selected faces, beginning
with one and ending with six. In Figure \ref{fig8} we present a graphical
display of the experimentally reported frequencies along with those
calculated theoretically by solving equation (\ref{8.5}). In Table 1 we
give the numerical values. As can be seen, coincidence between our
theoretical predictions and the experimental data is remarkable: the
worst error is 0.2\%, while for the most part it is below 0.1\%. Thus
{\it discrepancies between our theoretical predictions and experiment are
of order $\eta$\/}, as indeed expected ---see (\ref{8.7}). In addition, it
is also reported in reference \cite{jm97} that the 11-th, weakly coupled
mode of the {\sl TIGA\/} (enclosed in square brackets in Table 1) has a
practically zero amplitude, again in excellent agreement with our general
theoretical predictions about modes beyond the tenth ---see paragraph
after equation~(\ref{8.5}).

This is an encouraging result which motivated us to try a better fit by
estimating the {\it next order\/} corrections, i.e., $\chi_1$ of
(\ref{4.11.a}). As it turned out, however, matching between theory and
experiment did not consistently improve. This is not really that surprising,
though, as M\&J explicitly state \cite{jm97} that control of the general
experimental conditions in which data were obtained had a certain degree of
tolerance, and they actually show satisfaction that $\sim$1\% coincidence
between theory and measurement is comfortably accomplished. But 1\% is
{\it two orders of magnitude larger than $\eta\/$} ---cf. equation
(\ref{8.7})---, so failure to refine our frequency estimates to order
$\eta\/$ is again fully consistent with the accuracy of available real data.

A word on a technical issue is in order. Merkowitz and Johnson's equations
for the {\sl TIGA\/} \cite{jm93,jm95} are identical to ours to lowest order
in $\eta\/$. Remarkably, though, their reported theoretical estimates of
the system frequencies are not quite as accurate as ours \cite{jm97}.
The reason for this is probably the following: in M\&J's model these
frequencies appear within an algebraic system of 5+$J\/$ linear equations
with as many unknowns which has to be solved; in our model the algebraic
system has only $J\/$ equations and unknowns, actually equations
(\ref{3.16}). This is a very appreciable difference for the range
of values of $J\/$ under consideration. While the roots for the frequencies
can be seen to {\it mathematically\/} coincide in both approaches, in
actual practice these roots are {\it estimated\/}, generally by means of
computer programmes. It is here that problems most likely arise, for the
numerical reliability of an algorithm to solve matrix equations normally
decreases as the rank of the matrix increases. The significant algebraic
simplification of our model's equations should therefore be considered
one of real practical value.

\subsection{Other mismatched parameters}

We now assess the system sensitivity to small mismatches in resonators'
masses, locations and frequencies.

\subsubsection{Resonator mass mismatches}

If the {\it masses\/} are slightly non-equal then we can write

\begin{equation}
  M_a = \eta{\cal M}\,(1+\mu_a\,\eta^{1/2})\ ,\qquad a=1,\ldots,J
  \label{8.9}
\end{equation}
where $\eta\/$ can be defined e.g. as the ratio of the {\it average\/}
resonator mass to the sphere's mass. It is immediately obvious from
equation (\ref{8.9}) that mass non-uniformities of the resonators only
affect our equations in {\it second order\/}, since resonator mass
non-uniformities result, as we see, in corrections of order $\eta^{1/2}$
to $\eta^{1/2}$ itself, which is the very parameter of the perturbative
expansions. The system is thus clearly {\it robust\/} to mismatches in
the resonator masses of the type (\ref{8.9}).

\subsubsection{Errors in resonator locations}

The same happens if the {\it locations\/} of the resonators have tolerances
relative to a {\it pre-selected\/} distribution. For let ${\bf n}_a\/$ be a
set of resonator locations, for example the {\sl TIGA\/} or the {\sl PHC\/}
positions, and let ${\bf n'}_a\/$ be the real ones, close to the former:

\begin{equation}
  {\bf n'}_a = {\bf n}_a + {\bf v}_a\,\eta^{1/2}\ ,\qquad a=1,\ldots,J
  \label{8.10}
\end{equation}

The values ${\bf n}_a$ determine the eigenvalues $\zeta_a\/$ in equation
(\ref{5.2}), and also they appear as arguments to the spherical harmonics
in the system response functions of sections~\ref{sec:srgw}--\ref{sec:hs}.
It follows from~(\ref{8.10}) by continuity arguments that

\begin{eqnarray}
  Y_{lm}({\bf n'}_a) & = & Y_{lm}({\bf n}_a) + O(\eta^{1/2})
  \label{8.105.a} \\
  \zeta'_a & = & \zeta_a + O(\eta^{1/2})
  \label{8.105.b}
\end{eqnarray}

Inspection of the equations of sections~\ref{sec:srgw}--\ref{sec:hs} shows
that both $\zeta_a\/$ and $Y_{lm}({\bf n}_a)$ {\it always\/} appear within
lowest order terms, and hence that corrections to them of the
type~(\ref{8.105.a})-(\ref{8.105.b}) will affect those terms in {\it second
order\/} again. We thus conclude that the system is also {\it robust\/} to
small misalignments of the resonators relative to pre-established positions.

\subsubsection{Resonator frequency mistunings}

The resonator {\it frequencies\/} may also differ amongst them, so let

\begin{equation}
  \Omega_a = \Omega\,(1+\rho_a\,\eta^{1/2})\ ,\qquad a=1,\ldots,J
  \label{8.11}
\end{equation}

To assess the consequences of this, however, we must go back to equation
(\ref{3.18}) and see what the coefficients in its series solutions of the
type (\ref{4.11.a}) are. The procedure is very similar to that of section
\ref{s8.1}, and will not be repeated here; the lowest order coefficient
$\chi_\frac{1}{2}$ is seen to satisfy the algebraic equation

\begin{equation}
  \det\left[\delta_{ab} - \frac{1}{\chi_\frac{1}{2}}\,\sum_{c=0}^{J}\,
  \frac{\chi_{ac}^{(n_0l_0)}\,\delta_{cb}}{\chi_\frac{1}{2}-\rho_c}\right]
  = 0	  \label{8.12}
\end{equation}
which reduces to (\ref{5.1}) when all the $\rho\/$'s vanish, as expected.
This appears to potentially have significant effects on our results to
lowest order in $\eta^{1/2}$, but a more careful consideration of the facts
shows that it is probably unrealistic to think of such large tolerances in
resonator manufacturing as implied by equation (\ref{8.11}) in the first
place. In the {\sl TIGA\/} experiment, for example \cite{phd}, an error of
order $\eta^{1/2}$ would amount to around 50 Hz of mistuning between
resonators, an absurd figure by all means. In a full scale sphere
($\sim$40 tons, $\sim$3 metres in diameter, $\sim$800 Hz fundamental
quadrupole frequency, $\eta\/$\,$\sim$\,10$^{-5}$) the same error would
amount to between 5 Hz and 10 Hz in resonator mistunings for the lowest
frequency. This is probably excessive for a capacitive transducer, but may
be realistic for an inductive one. With this exception, it is thus more
appropriate to consider that resonator mistunings are at least of order
$\eta\/$. If this is the case, though, we see once more that the system
is quite insensitive to such mistunings.

Summing up the results of this section, we can say that the resonator
system dynamics is quite {\it robust\/} to small (of order $\eta^{1/2}$)
changes in its various parameters. The important exception is of course
the effect of suspension drillings, which do result in significant changes
relative to the ideally perfect device, but which can be relatively easily
calculated. This theoretical picture is fully supported by experiment, as
{\it robustness\/} in the parameters here considered has been reported in
reference \cite{jm97}.

\section{Conclusions}

A spherical GW antenna is a natural multimode device with very rich
potential capabilities to detect GWs on earth. But such detector is not
just a bare sphere, it requires a set of {\it motion sensors\/} to be
practically useful. It appears that transducers of the {\it resonant\/}
type are the best suited ones for an efficient performance of the detector.
Resonators however significantly interact with the sphere, and they affect
in particular its frequency spectrum and vibration modes in a specific
fashion, which has to be properly understood before reliable conclusions
can be drawn from the system readout.

The main objective of this paper has been the construction and development
of an elaborate theoretical model to describe the joint dynamics of a solid
elastic sphere and a set of {\it radial motion\/} resonators attached to
its surface at arbitrary locations, with the purpose to make predictions
of the system characteristics and response, in principle with arbitrary
mathematical precision.

We have shown that the solutions to our equations of motion are expressible
as an ascending series in powers of the small ``coupling constant''
$\eta\/$, the ratio of the average resonator mass to the mass of the large
sphere. The {\it lowest order\/} approximation corresponds to terms of order
$\eta^{1/2}$ and, to this order, we recover, and widely generalise, other
authors' results \cite{jm97,ts,grg}, obtained by them on the basis of
certain simplifying assumptions. This has in particular enabled us to assess
the system response for arbitrary resonator layouts, and to search the
equations for configurations other than the highly symmetric {\sl TIGA\/}.
This search has led us to make a specific proposal, the {\sl PHC\/}, which
is based on a pentagonally symmetric set of 5 rather than 6 resonators per
quadruopole mode sensed. The {\sl PHC\/} distribution has the very interesting
property that {\it mode channels\/} can be constructed from the resonators'
readouts, much in the same way as in the {\it TIGA\/} \cite{jm95}. In the
{\sl PHC\/} however a new and distinctive characteristic is present: different
{\it wave amplitudes\/} selectively couple to different {\it detector modes\/}
having different frequencies, so that the antenna's mode channels come at
different rather than equal frequencies. The {\sl PHC\/} philosophy can be
extended to make a {\it multifrequency\/} system by using resonators tuned
to the first two quadrupole harmonics of the sphere {\it and\/} to the first
monopole, an altogether 11 transducer set \cite{ls}.

The assessment of {\it symmetry failure\/} effects, as well as other
parameter departures form ideality, has also interested us here. This is
seen to receive a particularly clear treatment in our general scheme: the
theory transparently shows that the system is {\it robust\/} against
relative disturbances of order $\eta\/$ or smaller in any system parameters,
and provides a systematic procedure to assess larger tolerances ---up to
order $\eta^{1/2}$. The system is shown to still be robust to tolerances of
this order in some of its parameters, whilst it is not to others. Included
in the latter group is the effect of spherical symmetry breaking due to
system suspension in the laboratory, which causes {\it degeneracy lifting\/}
of the sphere's eigenfrequencies, now split up into multiplets. By using
our algorithms we have succeeded in numerically reproducing the reportedly
measured frequencies of the {\sl LSU\/} prototype antenna \cite{phd} with
a fully satisfactory precision of four decimal places. The experimentally
reported robustness of the system to resonator mislocations \cite{jm97} is
also in full agreement with our theoretical predictions.

The perturbative approach we have adopted is naturally open to refined
analysis of the system response in higher orders in $\eta\/$. For example,
we can systematically address the weaker coupling of non-quadrupole modes. 
It appears however that such refinements will be largely masked by
{\it noise\/} in a real system, as shown by Merkowitz and Johnson \cite{jm98},
and this must therefore be considered first. So our next step is to include
noise in the model and see its effect. Stevenson \cite{ts} has already made
some progress in this direction, and partly assessed the characteristics of
{\sl TIGA\/} and {\sl PHC\/}, but more needs to be done since not too high
signal-to-noise ratios should realistically be be considered in an actual
GW detector. In particular, the discovery of {\it mode channels\/} also for
the {\sl PHC\/} distribution opens the possibility of analysis of noise
correlations and dependencies, as well as the errors in GW parameter
estimation. These are natural extensions of this research, and some of them
are currently underway \cite{lms}.

\section*{Acknowledgments}

We are indebted with Stephen Merkowitz for his kind supply of the
{\sl TIGA\/} prototype data, without which a significant part of this
work would have been speculative. Fruitful discussions with him are also
gratefully acknowledged. We thank Eugenio Coccia for interaction and
encouragement throughout the development of this research, and also Curt
Cutler for pointing out to us an initial error in equation~(\ref{2.1.b}).
We have received financial support from the Spanish Ministry of Education
through contract number PB96-0384, and from the Institut d'Estudis
Catalans.

\appendix

\section{Green functions for the multiple resonator system}
\label{app:a}

The density of forces in he rhs of equation (\ref{2.1.a}) happens to be
of the {\it separable\/} type

\begin{equation}
  {\bf f}({\bf x},t) = \sum_\alpha\,{\bf f}^{(\alpha)}({\bf x})\,
   g^{(\alpha)}(t)   \label{3.1}
\end{equation}
where $\alpha\/$ is a suitable label. We recall from reference \cite{lobo}
that, in such circumstances, a formal solution can be written down for
equation (\ref{2.1.a}) in terms of a {\it Green function integral\/},
whereby the following orthogonal series expansion obtains:

\begin{equation}
   {\bf u}({\bf x},t) = \sum_\alpha\sum_N\,\omega_N^{-1}\,f_N^{(\alpha)}\,
   {\bf u}_N({\bf x})\,g_N^{(\alpha)}(t)      \label{3.2}
\end{equation}
where

\begin{eqnarray}
  f_N^{(\alpha)} & \equiv & \frac{1}{\cal M}\,\int_{\rm Sphere}
  {\bf u}_{N}^*({\bf x})\cdot{\bf f}^{(\alpha)}({\bf x})\,d^3x
  \label{3.3.a} \\[0.5 em]
  g_N^{(\alpha)}(t) & \equiv & \int_0^t g^{(\alpha)}(t')\,\sin\omega_N (t-t')
  \,dt'   \label{3.3.b}
\end{eqnarray}

Here, $\omega_N\/$ and ${\bf u}_N({\bf x})$ are the eigenfrequencies and
associated normalised wavefunctions of the free sphere. Also, $N\/$ is an
abbreviation for a multiple index $\{nlm\}$. The generic index $\alpha$
is a label for the different pieces of interaction happening in the system.
We quote the result of the calculations of the terms required in this paper:

\begin{eqnarray}
 f_{{\rm resonators,} N}^{(a)} & = & \frac{M_a}{\cal M}\,\Omega_a^2\,\,
  \left[{\bf n}_a\!\cdot\!{\bf u}_N^*({\bf x}_a)\right]\ ,\qquad a=1,\ldots,J
  \label{3.4.a} \\[0.5 em]
 f_{{\rm GW,}N}^{(l'm')} & = & a_{nl}\,\delta_{ll'}\,\delta_{mm'}\ \ ,
  \qquad N\equiv\{nlm\}\ ,\ \ l'=0,2\ ,\ \ m'=-l',\ldots,l'
  \label{3.4.b} \\[0.5 em]
 f_{{\rm stroke,}N} & = &{\cal M}^{-1}\,
                        {\bf f}_0\!\cdot\!{\bf u}_N^*({\bf x}_0)
  \label{3.4.c}
\end{eqnarray}
where the coefficients $a_{nl}\/$ in (\ref{3.4.b}) are overlapping integrals
of the type~(\ref{3.3.a}) \footnote{
We use here the better adapted notation $a_{n0}$ and $a_{n2}$ instead of
$a_n\/$ and $b_n\/$, respectively, of reference \protect\cite{lobo}.
Numerical values are however identical.}, and

\begin{eqnarray}
 g_{{\rm resonators,} N}^{(a)}(t) & = & \int_0^t\left[z_a(t')-u_a(t)
   \right]\,\sin\omega_N(t-t')\,dt'
   \ \ ,\qquad a=1,\ldots,J  \label{3.5.a}  \\[0.5 em]
 g_{{\rm GW,}N}^{(lm)}(t) & = & \int_0^t g^{(lm)}(t')\,
   \sin\omega_N(t-t')\,dt'  \label{3.5.b}  \\[0.7 em]
 g_{{\rm stroke,}N}(t) & = & \sin\omega_Nt  \label{3.5.c}
\end{eqnarray}

If this is replaced into (\ref{2.1.a}) one readily finds

\begin{equation}
 {\bf u}({\bf x},t) = \sum_N\,\omega_N^{-1}\,{\bf u}_N({\bf x})\,
  \left\{\sum_{b=1}^J\,\frac{M_b}{\cal M}\,\Omega_b^2\,
  \left[{\bf n}_b\!\cdot\!{\bf u}_N^*({\bf x}_b)\right]\,
  g_{{\rm resonators,} N}^{(b)}(t) + \sum_\alpha
  f_{{\rm external,} N}^{(\alpha)}\,g_{{\rm external,} N}^{(\alpha)}(t)
  \right\}	\label{3.6}
\end{equation}
where the label ``external'' explicitly refers to agents acting upon the
system from outside; two kinds of such external actions are considered in
this article: those due to GWs and those due to a calibration hammer stroke.
Specifying {\bf x}\,=\,${\bf x}_a\/$ in and multiply by ${\bf n}_a\/$ above,
the following is readily found:

\begin{eqnarray}
  u_a(t) & = & u_a^{\rm external}(t) + \sum_{b=1}^J\,\eta_b\,\int_0^t
  K_{ab}(t-t')\,\left[\,z_b(t')-u_b(t')\right]\,dt'  \label{A3.7.a}\\
  \ddot{z}_a(t) & = & \xi_a^{\rm external}(t)
  -\Omega_a^2\,\left[\,z_a(t)-u_a(t)\right]\ , \qquad a=1,\ldots,J
  \label{A3.7.b}
\end{eqnarray}
where $\eta_b$\,$\equiv$\,$M_b/{\cal M}$, $u_a^{\rm external}(t)$\,$\equiv$\,
${\bf n}_a\!\cdot\!{\bf u}^{\rm external}({\bf x}_a,t)$,

\begin{equation}
   {\bf u}^{\rm external}({\bf x},t) = \sum_\alpha\sum_N\,\omega_N^{-1}\,
    f_{{\rm external,}N}^{(\alpha)}\,{\bf u}_N({\bf x})\,
    g_{{\rm external,}N}^{(\alpha)}(t)      \label{3.9}
\end{equation}
and

\begin{equation}
  K_{ab}(t) \equiv \Omega_b^2\,\sum_N\,\omega_N^{-1}\,
  \left[{\bf n}_b\!\cdot\!{\bf u}_N^*({\bf x}_b)\right]
  \left[{\bf n}_a\!\cdot\!{\bf u}_N({\bf x}_a)\right]\,\sin\omega_Nt
  \label{A3.10}
\end{equation}

The following bare sphere responses to GWs and hammer strokes (equations
(\ref{1.1}) and (\ref{2.5}), respectively) can be calculated by direct
substitution. We present here the results as Laplace transform domain
functions, which are the useful ones for our purposes:

\begin{eqnarray}
  \hat u_a^{\rm GW}(s) & = &
   \sum_{\stackrel{\scriptstyle l=0\ {\rm and}\ 2}{m=-l,...,l}}\,\left(
   \sum_{n=1}^\infty\,\frac{a_{nl}\,A_{nl}(R)}{s^2+\omega_{nl}^2}\right)
   \,Y_{lm}({\bf n}_a)\,\hat g^{(lm)}(s)\ ,\qquad a=1,\ldots,J
   \label{6.2.a}	\\[1 em]
  \hat u_a^{\rm stroke}(s) & = & -\sum_{nl}\,\frac{f_0}{s^2+\omega_{nl}^2}\,
  \left|A_{nl}(R)\right|^2\,P_l({\bf n}_a\!\cdot\!{\bf n}_0)\ ,\qquad
   a=1,\ldots,J
  \label{6.2.b}
\end{eqnarray}
where $Y_{lm}\/$ are spherical harmonics and $P_l\/$ Legendre polynomials
\cite{Ed60}. The calculation of the Laplace transform of the kernel
matrix~(\ref{A3.10}) is likewise immediate:

\begin{equation}
  \hat K_{ab}(s) = \sum_N\,\frac{\Omega_b^2}{s^2+\omega_N^2}\,
   \left[{\bf n}_b\!\cdot\!{\bf u}_N^*({\bf x}_b)\right]
   \left[{\bf n}_a\!\cdot\!{\bf u}_N({\bf x}_a)\right]
   \label{3.17}
\end{equation}

Given that (see \cite{lobo} for full details)

\begin{equation}
   {\bf u}_{nlm}({\bf x}) = A_{nl}(r)\,Y_{lm}(\theta,\varphi)\,{\bf n}
   - B_{nl}(r)\,i{\bf n}\!\times\!{\bf L}Y_{lm}(\theta,\varphi)
   \label{A3.18}
\end{equation}
and that the spheroidal frequencies $\omega_{nl}\/$ are 2$l\/$+1--fold
degenerate, (\ref{3.17}) can be easily summed over the degeneracy index
$m\/$, to obtain

\begin{equation}
  \hat K_{ab}(s) = \sum_{nl}\,\frac{\Omega_b^2}{s^2+\omega_{nl}^2}\,
   \left|A_{nl}(R)\right|^2\,\left[\sum_{m=-l}^l\,
    Y_{lm}^*({\bf n}_b)\,Y_{lm}({\bf n}_a)\right]
   \label{A3.19}
\end{equation}
or, equivalently,

\begin{equation}
  \hat K_{ab}(s) = \sum_{nl}\,\frac{\Omega_b^2}{s^2+\omega_{nl}^2}\,
   \left|A_{nl}(R)\right|^2\,\frac{2l+1}{4\pi}\,
   P_l({\bf n}_a\!\cdot\!{\bf n}_b)
   \label{A3.20}
\end{equation}
where use has been made of the summation formula for the spherical
harmonics \cite{Ed60}

\begin{equation}
  \sum_{m=-l}^l\,Y_{lm}^*({\bf n}_b)\,Y_{lm}({\bf n}_a) = 
  \frac{2l+1}{4\pi}\,P_l({\bf n}_a\!\cdot\!{\bf n}_b)
  \label{A3.21}
\end{equation}
and where $P_l\/$ is a Legendre polynomial:

\begin{equation}
  P_l(z) = \frac{1}{2^l\,l!}\,\frac{d^l}{dz^l}\,(z^2-1)^l
  \label{A3.22}
\end{equation}

\section{System response algebra}
\label{app:b}

From equation (\ref{4.8}), i.e.,

\begin{equation}
 \sum_{b=1}^J\,\left[\delta_{ab} + \eta\,\sum_{nl}\,
   \frac{\Omega^2s^2}{(s^2+\Omega^2)(s^2+\omega_{nl}^2)}\,\chi_{ab}^{(nl)}
   \right]\,\hat q_b(s) = -\frac{s^2}{s^2+\Omega^2}\,
   \hat u_a^{\rm GW}(s) + \frac{\hat\xi_a^{\rm GW}(s)}
   {s^2+\Omega^2}\ ,\qquad  (\Omega = \omega_{n_0l_0})
   \label{A2.1}
\end{equation}
we must first isolate $\hat q_b(s)$, then find inverse Laplace transforms
to revert to time domain quantities. Substituting the values of
$\hat u_a^{\rm GW}(s)$ and $\hat\xi_a^{\rm GW}(s)$ from~(\ref{6.2.a})
and~(\ref{4.85}) into~(\ref{A2.1}) we find

\begin{equation}
   \hat q_a(s) = \sum_{\mbox{\scriptsize $\begin{array}{c}
    l=0\ \mbox{and}\ 2 \\ m=-l,...,l \end{array}$}}\hat\Phi_a^{(lm)}(s)\,
    \hat g^{(lm)}(s)\ ,\qquad a=1,\ldots,J
    \label{6.3}
\end{equation}
where

\begin{equation}
   \hat\Phi_a^{(lm)}(s) = -\frac{s^2}{s^2+\Omega^2}\,\left(-\frac{R}{s^2} +
   \sum_{n=1}^\infty\,\frac{a_{nl}\,A_{nl}(R)}{s^2+\omega_{nl}^2}\right)\,
   \sum_{b=1}^J\,\left[\delta_{ab} +
   \eta\,\sum_{nl}\,\frac{\Omega^2s^2}{(s^2+\Omega^2)(s^2+\omega_{nl})^2}\,
   \chi_{ab}^{(nl)}\right]^{-1}\,Y_{lm}({\bf n}_b)
   \label{6.4}
\end{equation}

Now, using the convolution theorem of Laplace transforms, we see that the
time domain version of equation~(\ref{6.3}) is

\begin{equation}
   q_a(t) = \sum_{lm}\,\int_0^t\,\Phi_a^{(lm)}(t-t')\,g^{(lm)}(t')\,dt'
   \ ,\qquad a=1,\ldots,J
   \label{6.5}
\end{equation}
where $\Phi_a^{(lm)}(t)$ is the {\it inverse Laplace transform\/} of
(\ref{6.4}). The inverse Laplace transform of $\hat\Phi_a^{(lm)}(s)$
can be expediently calculated by the {\it residue theorem\/} through
the formula \cite{hels}

\begin{equation}
   \Phi_a^{(lm)}(t) = 2\pi i\,\sum\,\left\{ {\rm residues\ of}\ \ 
   \left[\hat\Phi_a^{(lm)}(s)\,e^{st}\right]\ \ {\rm at\ its\ poles\ 
   in\ complex\ {\it s\/}-plane}\right\}
   \label{6.6}
\end{equation}

Clearly thus, the {\it poles\/} of $\hat\Phi_a^{(lm)}(s)$ must be determined
in the first place. It is immediately clear from equation~(\ref{6.4}) that
there are no poles at either $s\/$\,=\,0, or $s\/$\,=\,$\pm i\Omega$, or
$s\/$\,=\,$\pm i\omega_{nl}$, for there are exactly compensated infinities
at these locations. The only possible poles lie at those values of $s\/$
for which the matrix in square brackets in~(\ref{6.4}) is not invertible,
and these correspond to the zeroes of its determinant, i.e.,

\begin{equation}
  \Delta(s)\equiv\det\left[\delta_{ab} + \eta\,
  \frac{s^2}{s^2+\Omega^2}\,\hat K_{ab}(s)\right]= 0\ ,
  \qquad {\rm poles}	\label{3.18}
\end{equation}

There are infinitely many roots of equation (\ref{3.18}), and {\it analytic\/}
expressions cannot be found for them. {\it Perturbative\/} approximations in
terms of the small parameter $\eta\/$ will be applied instead. It is assumed
that

\begin{equation}
  \Omega = \omega_{n_0l_0}
\end{equation}
for a {\it fixed\/} multipole harmonic $\{n_0l_0\}$. Equation (\ref{3.18})
can then be recast in the more convenient form

\begin{equation}
  \Delta(s) \equiv \det\,\left[\delta_{ab} + \eta\,\frac{\Omega^2s^2}
   {(s^2+\Omega^2)^2}\,\chi_{ab}^{(n_0l_0)} + \eta\,\sum_{nl\neq n_0l_0}\,
   \frac{\Omega^2s^2}{(s^2+\Omega^2)(s^2+\omega_{nl}^2)}\,\chi_{ab}^{(nl)}
   \right] = 0
   \label{4.9}
\end{equation}

Since $\eta\/$ is a small parameter, the {\it denominators\/} of the
fractions in the different terms in square brackets in~(\ref{4.9}) must be
{\it quantities of order $\eta\/$} at the root locations for the determinant
to vanish at them. A distinction however arises depending on whether $s^2$ is
close to $-\Omega^2$ or to the other $-\omega_{nl}^2$. There are accordingly
two categories of roots, more precisely:

\begin{eqnarray}
    s_0^2 & = & -\Omega^2\,\left(1 + \chi_\frac{1}{2}\,\eta^{1/2}
    + \chi_1\,\eta + \ldots\right)  \qquad (\Omega=\omega_{n_0l_0})
    \label{4.11.a}   \\
    s_{nl}^2 & =& -\omega_{nl}^2\,\left(1 + b_1^{(nl)}\,\eta +
	b_2^{(nl)}\,\eta^2 + \ldots\right) \qquad ({nl\neq n_0l_0})
    \label{4.11.b}
\end{eqnarray}

The coefficients $\chi_\frac{1}{2}$, $\chi_1$,... and $b_1^{(nl)}$,
$b_1^{(nl)}$,... can be calculated recursively, starting form the first, by
substitution of the corresponding series expansions into equation~(\ref{4.9}).
The lowest order terms are easily seen to be given by

\begin{equation}
  \det\left[\delta_{ab} - \frac{1}{\chi_\frac{1}{2}^2}\,
  \chi_{ab}^{(n_0l_0)}\right] = 0
  \label{5.1}
\end{equation}
and

\begin{equation}
  \det\left[\frac{\Omega^2-\omega_{nl}^2}{\omega_{nl}^2}\,b_1^{(nl)}\,
  \delta_{ab} - \chi_{ab}^{(nl)}\right] = 0
  \label{5.4}
\end{equation}
respectively. Both equations (\ref{5.1}) and (\ref{5.4}) are algebraic
eigenvalue equations. As shown in appendix~\ref{app:c}, the matrix
$\chi_{ab}^{(nl)}$ has at most $(2l+1)$ non-null positive eigenvalues
---all the rest up to $J\/$ are identically zero.

As a final step we must evaluate (\ref{6.6}). This is accomplished by
standard textbook techniques (see e.g.\ \cite{porter}); the algebra is
quite straightforward but rather lengthy, and we shall not delve into its
details here, but quote only the most interesting results. It appears that
the {\it dominant\/} contribution to $\Phi_a^{(lm)}(t)$ comes from the poles
at the locations (\ref{4.11.a}), whereas all other poles only contribute as
higher order corrections; generically, $\Phi_a^{(lm)}(t)$ is seen to have
the form

\begin{equation}
   \Phi_a^{(lm)}(t)\propto\eta^{-1/2}\,\sum_{\zeta_c\neq 0}\,
   \left(\sin\omega_{c+}t - \sin\omega_{c-}t\right)\,\delta_{ll_0}
   + O(0)	\label{6.7}
\end{equation}
where

\begin{equation}
  \omega_{a\pm}^2 = \Omega^2\,\left(1\pm\sqrt{\frac{2l+1}{4\pi}}\,
  \left|A_{n_0l_0}(R)\right|\,\zeta_a\,\eta^{1/2}\right) + O(\eta)\ ,
  \qquad a=1,\ldots,J
\end{equation}

In Laplace domain one has,

\begin{equation}
   \hat\Phi_a^{(lm)}(s)\propto\eta^{-1/2}\,\sum_{\zeta_c\neq 0}\,
   \left[\left(s^2+\omega_{c+}^2\right)^{-1} -
   \left(s^2+\omega_{c-}^2\right)^{-1}\right]\,\delta_{ll_0}
   \label{6.78}
\end{equation}

Detailed calculation of the residues \cite{serrano} yield equation
(\ref{6.8}), which must be evaluated for each particular tuning and
resonator distribution, as described in section~\ref{sec:srgw}.

\section{Eigenvalue properties}
\label{app:c}

We give in this Appendix a few important properties of the matrix
$P_l({\bf n}_a\!\cdot\!{\bf n}_b)$ for arbitrary $l\/$ and resonator
locations ${\bf n}_a\/$ ($a\/$=1,\ldots,$J\/$) which are useful for
detailed system resonance characterisation.

We first note the {\it summation formula\/} for spherical harmonics
\cite{Ed60}

\begin{equation}
  \sum_{m=-l}^l\,Y_{lm}^*({\bf n}_a)\,Y_{lm}({\bf n}_b) = 
  \frac{2l+1}{4\pi}\,P_l({\bf n}_a\!\cdot\!{\bf n}_b)\ ,\qquad
  a,b=1,\ldots,J        \label{A.1}
\end{equation}
where $P_l\/$ is a Legendre polynomial:

\begin{equation}
  P_l(z) = \frac{1}{2^l\,l!}\,\frac{d^l}{dz^l}\,(z^2-1)^l   \label{A.15}
\end{equation}

To ease the notation we shall use the symbol ${\cal P}_l\/$ to mean the
entire $J\/$$\times$$J\/$ matrix $P_l({\bf n}_a\!\cdot\!{\bf n}_b)$, and
introduce Dirac {\it kets\/} $|m\rangle$ for the column $J\/$-vectors

\begin{equation}
  |m\rangle\equiv\sqrt{\frac{4\pi}{2l+1}}\left(\begin{array}{c}
    Y_{lm}({\bf n}_1) \\ \vdots  \\ Y_{lm}({\bf n}_J)
  \end{array}\right)\ \ ,\qquad m=-l,\ldots,l
  \label{A.2}
\end{equation}

These kets are {\it not\/} normalised; in terms of them equation (\ref{A.1})
can be rewritten in the more compact form

\begin{equation}
  {\cal P}_l = \sum_{m=-l}^l\,|m\rangle\langle m|     \label{A.3}
\end{equation}

Equation (\ref{A.3}) indicates that the {\it rank\/} of the matrix
${\cal P}_l\/$ cannot exceed $(2l+1)$, as there are only $(2l+1)$ kets
$|m\rangle$. So, if $J\/$\,$>$\,$(2l+1)$ then it has at least
$(J-2l-1)$ identically null eigenvalues ---there can be more if some of
the ${\bf n}_a\/$'s are parallel, as this causes rows (or columns) of
${\cal P}_l\/$ to be repeated.

We now prove that the non-null eigenvalues of ${\cal P}_l\/$ are
{\it positive\/}. Clearly, a regular eigenvector, $|\phi\rangle$, say, of
${\cal P}_l\/$ will be a linear combination of the kets $|m\rangle$:

\begin{equation}
  {\cal P}_l \,|\phi\rangle = \zeta^2\,|\phi\rangle\ ,\qquad
  |\phi\rangle = \sum_{m=-l}^l\,\phi_m\,|m\rangle
  \label{A.4}
\end{equation}
where we have called $\zeta^2$ the corresponding eigenvalue, since it is a
positive number, as we shall shortly prove. If the second (\ref{A.4}) is
substituted into the first then it is immediately seen that

\begin{equation}
  \sum_{m'=-l}^l\,\left(\zeta^2\,\delta_{mm'} -
  \langle m|m'\rangle\right)\,\phi_{m'} = 0    \label{A.5}
\end{equation}
which admits non-trivial solutions if and only if

\begin{equation}
  \det\,\left(\zeta^2\,\delta_{mm'} -
  \langle m|m'\rangle\right) = 0    \label{A.6}
\end{equation}

In other words, $\zeta^2$ are the eigenvalues of the
$(2l+1)$$\times$$(2l+1)$ matrix $\langle m|m'\rangle$, which is positive
definite because so is the ``scalar product'' $\langle\phi|\phi'\rangle$.
All of them are therefore strictly positive.

Finally, since the {\it trace\/} is an invariant property of a matrix, and

\begin{equation}
  {\rm trace}({\cal P}_l) \equiv \sum_{a=1}^J\,
  P_l({\bf n}_a\!\cdot\!{\bf n}_a) = \sum_{a=1}^J\,1 = J
  \label{A.7}
\end{equation}
we see that the eigenvalues $\zeta_a^2$ add up to $J\/$:

\begin{equation}
  {\rm trace}({\cal P}_l) =
  \sum_{a=1}^J\,\zeta_a^2\equiv\sum_{\zeta_a\neq 0}\,\zeta_a^2 = J
\end{equation}

\bsp

\label{lastpage}

\end{document}